\newif\iffinal
\finalfalse
\iffinal
\documentclass[]{IEEEtran} %
\else
\documentclass[11pt,onecolumn,draftcls]{IEEEtran} %
\fi

\usepackage{multirow}
\usepackage{amsfonts,amssymb,amsmath,amsthm,mathrsfs}
\usepackage{algorithmic,algorithm}
\usepackage{xcolor}
\usepackage{ifpdf}
\ifpdf
\usepackage[]{graphicx}
\else
\usepackage[draft]{graphicx}
\fi
\usepackage{array}
\usepackage{float}
\usepackage{enumitem}
\usepackage[caption = false]{subfig}
\usepackage{caption}
\usepackage{url,cite}
\usepackage{breakurl}
\usepackage{changepage}

\def\Pr{\mathsf{Pr}}
\def\Pub{\mathsf{Pub}}
\def\CC{\mathsf{CC}}

\def\SHA{\mathsf{SHA256}}
\def\HMAC{\mathsf{HMAC}\mbox{-}\mathsf{SHA512}}
\def\leftbits{\mathsf{Left256}}
\def\rightbits{\mathsf{Right256}}
\def\bip{\mathsf{BIP}}
\def\RIPEMD{\mathsf{RIPEMD160}}
\def\id{\mathsf{ID}}
\def\utxo{\mathsf{UTXO}}
\def\tx{\mathsf{TX}}
\def\B{\mathsf{B}}
\def\M{\mathsf{M}}
\def\secp{\mathsf{secp256k1}}
\def\CSPRNG{\mathsf{CSPRNG}}
\def\version{\mathsf{version}}
\def\verack{\mathsf{verack}}
\def\addr{\mathsf{addr}}
\def\getaddr{\mathsf{getaddr}}
\def\inv{\mathsf{inv}}
\def\getdata{\mathsf{getdata}}
\def\txx{\mathsf{tx}}
\def\aes{\mathsf{AES}}
\def\Encrypt{\mathsf{Encrypt}}

\def\para#1{\smallskip\noindent\emph{#1}:\,}

\usepackage[normalem]{ulem}
\definecolor{revision}{rgb}{1,0,0}

\begin{document}
\title{Bitcoin and Blockchain: \textit{Security and Privacy}}
\author{Ehab Zaghloul, Tongtong Li, Matt Mutka and Jian Ren%
\thanks{The authors are with the Department of Electrical and Computer Engineering, Michigan State University, East Lansing, MI 48824-1226, Email: \{ebz, tongli, mutka, renjian\}@msu.edu}}

\maketitle

\begin{abstract}
A cryptocurrency is a decentralized digital currency that is designed for secure and private asset transfer and storage. 
As a currency, it should be difficult to counterfeit and double-spend. 
In this paper, we review and analyze the major security and privacy issues of Bitcoin. 
In particular, we focus on its underlying foundation, blockchain technology.
First, we present a comprehensive background of Bitcoin and the preliminary on security.
Second, the major security threats and countermeasures of Bitcoin are investigated. 
We analyze the risk of double-spending attacks, evaluate the probability of success in performing the attacks and derive the profitability for the attacker to perform such attacks. 
Third, we analyze the underlying Bitcoin peer-to-peer network security risks and Bitcoin storage security.
We compare three types of Bitcoin wallets in terms of security, type of services and their trade-offs.
Finally, we discuss the security and privacy features of alternative cryptocurrencies and present an overview of emerging technologies today.  Our results can help Bitcoin users to determine a trade-off between the risk of double-spending attempts and the transaction time delay or confidence before accepting transactions.
These results can also assist miners to develop suitable strategies to get involved in the mining process and maximize their profits. 
\end{abstract}


\section{Introduction}

A cryptocurrency is a decentralized online currency that was developed as an alternate means to transfer money in an unprecedented way.
Existing financial systems require a centralized trusted financial institution to securely process transactions between two parties.
This institution charges costly service fees that are unavoidable for banking customers. 
In addition to such cost burdens, delayed processing time and security issues have affected the modern-day financial industry.
Certain transactions, such as funds transfer, may take days or weeks to be cleared, causing issues in cases of urgency. 
The modern-day financial system is also plagued with security and privacy vulnerabilities.
Financial institutions employ the most advanced security techniques to protect customers. 
However, the sensitive information of the customer is always exposed to the financial institutions making it vulnerable to information leakage.
To mitigate these security concerns, privacy risks, and inconveniences, new cryptographic protocols have been developed to allow secure and convenient asset transfer, without involving a centralized third-party. 

In 2008, Satoshi Nakamoto developed a white paper in which he proposed Bitcoin~\cite{Nakamoto08Bitcoin}. 
Bitcoin is an online Peer-to-Peer (P2P) digital cash system that does not require a trusted third-party.
In Bitcoin, users possess ownership rights to virtual cryptocoins that are denoted as Bitcoins~(BTC).
Users generate transactions to transfer BTC and store them in the public ledger, blockchain.
The smallest transferable value today is known as a Satoshi, which is equivalent to one-hundredth of a millionth BTC (i.e. 0.00000001 BTC). 

Bitcoin transactions utilize cryptographic protocols to provide a secure process while striving to preserve the privacy of both the buyer and seller.
The transactions are stored in a blockchain~\cite{haber1990time,bayer1993improving,haber1997secure,massias1999design} to limit inherent issues of digital media such as double-spending~\cite{karame2012two}.
A blockchain is a distributed database acting as a public ledger that holds all processed transactions.
It is based on a distributed consensus that allows any past and present online transaction to be verified ~\cite{crosby2016blockchain}.

Bitcoin transactions are released into the network and validated by the nodes as they propagate through the entire network.
The validating nodes, referred to as \textit{miners}, compete to mine groups of transactions into blocks and earn BTC as a reward. Mining is the process of solving a hard cryptopuzzle, referred to as the \textit{Proof-of-Work} (PoW), that requires extensive computational power.
The first miner capable of finding a solution to the problem broadcasts his/her block to the network and earns the reward.
The reward consists of a specified amount of new released BTC and all the transaction fees associated with the transactions included in the block.
All the other miners then surrender to the solution of the winning miner and append the winning block to the blockchain.

The first Bitcoin software was implemented by Satoshi Nakamoto and is known as the Bitcoin Core. 
This implementation is sometimes referred to as the \textit{Satoshi client} and is run by most of the network nodes in Bitcoin.
It is an open source project with a large developer community contributing to it.
The developers follow a Bitcoin Improvement Proposals~(BIP)~\cite{bip2} document and introduce the standards of the system. 
The document also contains new features and proposals for the developer community to test thoroughly before making final modifications to the software.

Following Bitcoin, many cryptocurrency systems appeared and continue to do so today. 
The blockchain technology is a common characteristic shared by many newly emerging cryptocurrency systems~\cite{hileman2017global}.
The majority of these systems are mainly clones of Bitcoin. 
These systems introduced only minor adjustments such as currency supply or block size within the Blockchain.
Alternatively, a few systems introduce innovative concepts that offer substantial features. 
Examples of these features include novel consensus mechanisms or enhanced decentralized computing platforms that can provide additional functions and higher flexibility to the system.

All cryptocurrencies are traded in the online cryptocurrency marketplace. 
The cryptocurrency market is similar to other exchange markets such as the stock market, with various trading platforms. 
However, the cryptocurrency market is not regulated by a government or agency and trading occurs virtually 24/7 across the world. 
The nature of cryptocurrency allows transactions to occur at speeds that cannot be accomplished with fiat currency, such as the United States dollar. 
This results in a much more volatile market than traditional trading markets. 
Coin prices are continuously rising and dropping, and new cryptocurrencies consistently enter and leave the marketplace. 
Many coins continue to rise in value based on value demonstrated to investors. 
However, increased speculation in the marketplace has lead to the over-evaluation of many cryptocurrencies.

As of November  2017, the total market cap of the cryptocurrency market hit \$246 billion~\cite{coinmarketcap}. This amount comes from the total valuation of almost one thousand cryptocurrencies on the market today.
Comparing this amount to the \$17.6 billion total market cap in 2016, the market has increased by 1,298\%.
This rapid growth in the new market has led an effort to examine the role of cryptocurrency in the future.

\subsection{Contributions}
Cryptocurrencies, particularly Bitcoin, have attracted massive and diverse attention.
They are in continuous development and evolution thrusting researchers to thoroughly and constantly analyze them.
Notable studies have been presented that discuss the blockchain technology and outline open issues.
Their main purpose is to exploit the future stability of Bitcoin from different perspectives.
The study presented in~\cite{bonneau2015sok} is one of the first studies to present an exposition of Bitcoin and some altcoins.
This work focused on discussing stability properties and comparing them to those in Bitcoin, to measure its degree of stability as a system.
It also briefly investigated security and privacy concerns, in addition to some alternative consensus protocols.
However, the study lacks recently investigated attacks and deep analysis behind them.
It is also limited in its discussions about the alternative protocols.
The survey presented in~\cite{tschorsch2016bitcoin} is a technical analysis on Bitcoin and aimed at consolidating key algorithmic features of the system.
This work expanded on the study in~\cite{bonneau2015sok} by providing an in-depth analysis in terms of security and privacy.
However, the authors briefly discuss Bitcoin storage wallets and do not delve into the underlying infrastructure that secures these wallets.
The study presented in~\cite{conti2017survey} is another comprehensive survey that explored various security and privacy aspects of Bitcoin, accompanied with possible countermeasures.
The survey presents comparisons between various security attacks and privacy protocols.
Nevertheless, as in~\cite{tschorsch2016bitcoin}, Bitcoin storage wallet types are briefly compared without exploiting their foundation that reflects points of weakness.
Another privacy-focused research is presented in~\cite{khalilov2018survey} that aims at expanding the previous works in terms of anonymity and privacy.
In this study, the authors show that analyzing anonymity and privacy in Bitcoin may be classified into various classes where each class may result in a different privacy leakage outcomes.
They also classify the on-going efforts to improve anonymity and privacy while discussing their potential corresponding outcomes.

While the main purpose of this paper is similar to the previous studies; to examine the potential stability of Bitcoin, however, we strive to approach this goal by considering missing and limited previously discussed topics.
We delve deeper into analyzing the double-spending attacks by modeling the probability of success in multiple ways.
In particular, utilizing our analysis, we present our own profitability analysis of the double-spending attacks.
We reveal a break-point in time when attackers should give up on the attack since it is unlikely that they will turn a profit beyond this point (i.e. the time when the cost is greater than the revenue).
We present a trade-off between the waiting time before accepting a transaction versus the profits/losses of the attackers.
This may help maximize the confidence of the users before accepting transactions. 
In addition to this, we thoroughly analyze the infrastructure of Bitcoin storage wallets.
Our discussion presents the key algorithmic features introduced by each wallet type in order to counter the different potential threats.
The main purpose is to enlighten the users with the trade-offs when using different types of wallets from a cryptographic perspective.

The major contributions of this paper can be summarized as follows:
\begin{enumerate}

\item We provide a comprehensive explanation of the primary components of Bitcoin discussed in a sequential and logical order for the readers to comprehend.
The main purpose is to cultivate the readers with the necessary background on Bitcoin to consolidate their understanding of the system.
We aim at providing sufficient background for the readers to build a solid understanding that can be utilized when exploring similar systems.
This background will also help the readers easily digest the following concerns and issues discussed in this paper.

\item We delve thoroughly into the analysis of double-spending attacks. 
We first show that the probability of success of performing double-spending attacks can be modeled using two distinct probabilistic models.
We show that both models result in a similar outcome.
Next, using these probabilistic models, we present a profitability analysis on performing double-spending attacks.
The main purpose of this analysis is to reflect the trade-off between the waiting time before accepting a transaction versus the profits/losses of the attackers.
We also aim at reflecting that attackers with 51\% computational power or more will continue to profit indefinitely.

\item We present fundamental network security and privacy concerns. 
The purpose of this analysis is to expand the knowledge of readers on major security and privacy concerns that threaten the stability of systems running the blockchain technology.
Our target is to help the reader realize the major threats to such systems from a security and privacy perspective.

\item We dive deeply into the exploration of Bitcoin storage wallets.
We first classify wallets based on their underlying infrastructure and methods of PKI pair generation.
We aim at presenting the cryptographic primitives related to each type of wallet.
Next, we classify wallets based on installation environments and then further classify them based on functionality. 
We strive to help the readers understand the different classes of wallets, their corresponding security risks, and the best practices to secure their cryptocoins.
\end{enumerate}

The interest in blockchain continues to grow aggressively. 
It has already attracted a wide range of audiences such as governments, enterprises, health-care, and many more.
We realize that in order for blockchain to sustain its success and for these interested entities to adopt it, we must educate a wider range of audience, which could include: (i) researchers at the beginning of the line that wish to expand on research in this area and (ii) skeptical entities and individuals that wish to adopt the technology and wish to learn more about it.
We aim at putting together a comprehensive study that explores blockchain technology from multiple angles and filling in the gaps of previous studies.

\subsection{Organization}
The rest of the paper is organized as follows. 
In Section \ref{Sec:Related-Work}, we briefly review previous digital cash systems and blockchain infrastructures. 
In Section \ref{Sec:Bitcoin}, we provide a comprehensive background review on Bitcoin outlining its building blocks and protocols.
Next, in Section \ref{Sec:DS}, we evaluate double-spending attacks and present our profitability analysis.
Following that, in Section \ref{Sec:networksecurity}, we assess the major network security attacks of the Bitcoin network.
In Section \ref{Sec:storagesecurity}, we analyze the security issues in the storage wallets used by Bitcoin today.
We investigate the subsequent privacy protocols of Bitcoin in an effort to limit the linkage problem in Section \ref{Sec:privacy}.
In Section \ref{Sec:altcoins}, we review protocols and alternative consensus algorithms implemented in emerging cryptocurrencies outlining the security and privacy advantages and limitations.
Finally, we conclude our study, summarize the lessons learned, and future research directions in Section \ref{Sec:conc}.

\section{Related Work} \label{Sec:Related-Work}
In this section, we discuss the history of digital cash systems.
We also introduce the evolution of the blockchain technology.

\subsection{Digital Cash Systems}
Research in digital cash dates back to the early 1980s~\cite{chaum1983blind}.
In 1990, DigiCash Inc., an electronic cash corporation, made an initial attempt to provide a cryptocurrency system~\cite{chaum1990untraceable}.
DigiCash transactions involved cryptographic protocols and aimed at providing its users with anonymity. 
However,it failed in 2000 as the Internet bubble popped despite being attractive initially.
David Chaum, its founder,  believes the failure of DigiCash to succeed was tied to its technology which preceded the e-commerce maturation within the Internet. Other reasons which led to its failure included the cooperation of banks to process a transaction, making DigiCash a centralized system.

In 1998, a decentralized digital cash system, b-money~\cite{bmoney}, was introduced by Wei Dai.
B-money is an anonymous and distributed digital cash system that aimed at providing untraceable transactions.
One major advantage of this system is that it eliminates the need for a central authority.
However, b-money was just an initial and incomplete idea. 
It did not properly tackle some of the key issues including double-spending attacks.

In early 2000, Digital Gold Currency (DGC), a currency backed by gold, gained some popularity. 
DGC is considered to be a second-generation digital currency. 
It is issued by some companies that enable users to pay each other in units similar to those of gold bullion.
Examples include iGolder, gbullion, and e-Gold. 
Although DGC seemed to have a bright future, it lost popularity due to its centralized structure. 
Politics may have also played a role in its declining popularity. 
Companies that provided DGC were forced to shut down by the federal government due to their inability to comply with the government regulations~\cite{frunza2015solving}.

In 2003, Second Life~\cite{secondlife}, an online virtual world, introduced a digital currency referred to as the Linden dollar.
The Linden dollar is exchangeable for fiat currencies. 
Second Life users are able to use this currency for direct transactions. However, similar to its preceding digital currencies, the Linden dollar is a centralized digital currency that is controlled by its creator, Linden Labs~\cite{linden}. 
Moreover, its price is volatile and unstable making it a risky currency to own.

\subsection{Blockchain History}
In 1991, the first secure blockchain was proposed by Stuart Haber and W. Scott Stornetta~\cite{haber1990time}.
Their blockchain aimed at certifying the creation or modification of a digital record by digitally time-stamping the record being processed.
However, the blockchain was not efficient since each record was independently time-stamped.
To improve the efficiency, Merkle trees~\cite{merkle1987digital} were incorporated into blockchains in 1992~\cite{bayer1993improving}.
They improved the efficiency by handling multiple digital records into one block.
Finally, Satoshi Nakamoto implemented the first real blockchain and used it as the core technology for the Bitcoin cryptocurrency system.

\section{Understanding Bitcoin} \label{Sec:Bitcoin}
In this section, we will present the major building blocks and protocols of Bitcoin.
We first present the Bitcoin network, Bitcoin transaction, and Bitcoin transaction standards.
Next, we explain how Merkle trees are utilized to group transactions into blocks and stored in the blockchain.
Following that, we discuss the Bitcoin mining process, mining pools, payment methods, and methods of developing alternative cryptocurrencies.

\subsection{The Bitcoin Network}
Bitcoin runs over a P2P network.
The main advantage of using a P2P network is the agile movement of data for all nodes to achieve consensus.
In contrast to the typical P2P network used to share data files between interested peers, Bitcoin utilizes the network to rapidly broadcast data among all the connected nodes.
This process is known as \textit{flooding} and continues until all nodes within the network receive the broadcast data.

It is important to differentiate between the terms \textit{node} and \textit{peer} of a P2P network.
A node is a network entity that is connected to one or multiple other similar nodes.
The directly connected nodes are referred to as the peers.
Nodes propagate data to the indirectly connected nodes by traversing it to their peers which follow a similar manner until the data reaches every connected node.

In the Bitcoin network, data being flooded includes IP addresses of the nodes, newly generated transactions, and blocks of verified transactions that extend the blockchain.
Peers share IP addresses of other nodes that they are connected to or have discovered from their peer nodes.
The aim behind sharing IP addresses is to allow peers in the network to discover and connect to more nodes resulting in a random network topology.
Newly generated transactions are broadcast through the network to rapidly publicize their occurrence to all connected nodes.
Miners compete to mine these transactions into blocks. 
The winning miner broadcasts the block to all the connected nodes to extend and update their version of the blockchain.

Nodes in the Bitcoin P2P network are defined based on their roles.
The main duties are summarized as transaction generation, block/transaction routing, block/transaction verification, and transaction mining.
Block/transaction routing is performed by all nodes.

A node that can perform all functions is referred to as a \textit{full node}.
It consistently keeps a copy of the full blockchain allowing it to verify any transaction without needing assistance of other connected nodes.
It also possesses a BTC wallet that can generate transactions and compute the possessed value of BTC by the node.
Moreover, a full node possesses computational resources to compete in the mining competition.
Nodes that do not store a full copy of the blockchain are referred to as \textit{Simplified Payment Verification (SPV) nodes} or \textit{lightweight nodes}.
These nodes require assistance from full nodes when verifying a transaction.
Full nodes feed the SPV nodes with the required information from the blockchain necessary to complete the transaction verification.

Some nodes may only perform one particular function.
Ones that are engaged in the mining process are referred to as \textit{mining nodes} while others that generate transactions are referred to as \textit{wallets}.

In most Bitcoin software implementations, all nodes are treated equally and can be uniquely identified by their IP addresses.
Using these addresses, peers establish Transmission Control Protocol (TCP) connections with one another.
Each node can choose whether to connect to the network using a public or private IP.
A node that utilizes a public IP is accessible over the Internet by any connected node while one with a private IP is only accessible by nodes within its private network.
By default, a node with a public IP address is granted 8 \textit{outbound} connections and 117 \textit{inbound} connections, resulting in a total of 125 connections.
On the other hand, a node with a private IP address is granted only 8 \textit{outbound} connections. 
An outbound connection is initiated by the node itself when it requests connecting to a discoverable node while an inbound connection is initiated by other nodes in the network that desire connecting to the node.

For explanation purposes, we define the node that initiates a connection as \textit{client} and the node that waits for an incoming connection as \textit{server}.
Both nodes engage in a TCP handshake by exchanging network packets defined as $\version$ and $\verack$.
The client initiates a connection request by sending a $\version$ packet addressed to the IP address of the server.
By default, the server listens on port 8333 for incoming $\version$ packets.
If the server accepts the $\version$ packet, it responds with a $\verack$ packet and its own $\version$ packet, both addressed to the IP address of the client.
Finally, the client responds by sending a $\verack$ packet addressed to the IP address of the server and the connection is established.
The connection enables symmetric communication allowing the client and server to exchange data bidirectionally.
The connection is lost if peers do not communicate for a specified idle time.  
To reconnect, peers engage in a new TCP handshake.

As discussed previously, a node shares with its peers a list of IP addresses that it has learned as a result of being connected to the network.
Each node stores its list in two separate tables: a \textit{tried} table and a \textit{new} table.
The tried table of a node stores IP addresses that the node has established connections with while its new table stores IP addresses that it has only discovered but did not attempt to connect to yet.
When a node desires sharing IP addresses with its peers, it randomly selects IP addresses from both tables and sends them in $\addr$ messages.
An $\addr$ message can contain up to 23\% or a maximum of 1000 IP addresses of the total IP addresses stored in both tables.
To initiate sharing, a node sends a $\getaddr$ message to its peers requesting them to share their lists of IP addresses.
The peers then respond with an $\addr$ message.
In some cases, sharing IP addresses is unsolicited if a node voluntarily sends an $\addr$ messages to its peers without receiving a $\getaddr$ message.

A node that wishes to connect to the Bitcoin network for the first time cannot obtain IP addresses by this method.
Bootstrapping is mainly achieved by communicating with a Domain Name Server (DNS) seeder.
The node sends a DNS query requesting a list of active IP addresses.
If the DNS fails to respond with an appropriate list of active IP addresses, the node can still connect to the network by using a hard-coded list of IP addresses, referred to as \textit{seeds}.
Once connected to any of these IP addresses, the node can then request more IP addresses from its peers by sending $\getaddr$ messages.

Nodes also relay verified transactions and blocks to their peers to reach consensus.
A node begins by broadcasting an inventory ($\inv$) message to all its peers informing them of the new transactions or blocks it has received and verified.
The peer nodes check whether they are already informed of these new transactions and blocks then respond to the node with a $\getdata$ message.
The $\getdata$ message includes all the transactions and blocks a peer node is not aware of.
The node then responds with a transaction/block message that includes the complete transactions/blocks the peer requests.
Once received, the peer validates the transactions or blocks and continues to relay them to its own peers in a similar manner.
If a received transaction or block cannot be validated, it is immediately dropped and its propagation is discontinued.

\subsection{Bitcoin Transactions} \label{subsec:transaction}

We define a Bitcoin transaction ($\tx$) as the transfer of an amount of BTC \textit{ownership rights} from the wallet of the buyer to the wallet of the seller, in exchange for a product or service. 
BTC wallets utilize elliptic curve digital signatures to handle the transfer of ownership rights and ensure that unauthorized spending of the cryptocurrency is infeasible.  
Each wallet randomly generates a private key $\Pr$ which is used to derive its corresponding public key $\Pub$ that is shared among all users.
The $\Pub$ is used to generate the address of the wallet needed to make payments to it while the $\Pr$ is used to generate a digital signature corresponding to the $\Pub$ in order to claim payments made to the wallet and use them in later transactions.
A $\Pr$ is first generated from a Cryptographically Secure Pseudo-Random Number Generator~($\CSPRNG$) and its corresponding $\Pub$ is then calculated using Elliptic Curve Digital Signature Algorithm~(ECDSA).
Calculations are performed based on the field and curve parameters defined by $\secp$ with the curve order $n$~\cite{brown2010} as follows
\begin{align} 
\Pr &= \CSPRNG() \label{eq:privatekey} , \\
\Pub &= \Pr \times G  ~(\bmod~n)\label{eq:publickey},
\end{align}
where $G$ is a generator of the elliptic curve and $\times$ represents elliptic curve multiplication.

\begin{figure}[t]
\centering
\includegraphics[scale=1.1]{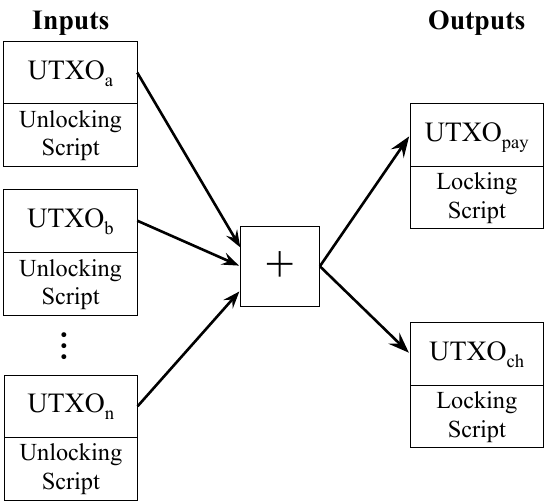}
\caption{A single transaction with multiple $\utxo$ inputs and outputs.}
\label{Fig:transaction}
\end{figure}

The BTC wallet of the buyer assembles a transaction using the Unspent Transaction Outputs ($\utxo$) of the buyer stored in the blockchain.
An $\utxo$ specifies an amount of BTC claimed earlier by the buyer as a result of a previously processed transaction.
A simple BTC transaction is shown in Fig. \ref{Fig:transaction}. 

In the figure, we show that a transaction can consist of multiple inputs and outputs. 
The output $\utxo_{pay}$ represents the transfer of ownership rights of a certain amount of BTC from the wallet of the buyer to the wallet of the seller.
The output $\utxo_{ch}$ represents redirecting ownership rights of the BTC change amount back to the wallet of the buyer.
A distinct \textit{locking script} is attached to each of these outputs which specifies conditions that must be met in order to grant ownership rights.
For example, the locking script attached to $\utxo_{pay}$ must include the $\Pub$ of the seller needed to generate his/her wallet address.
This ensures that the payment is made to the wallet of the seller and only he/she is granted access to it with his/her corresponding $\Pr$.
Using $\Pr$, the seller can generate a digital signature that corresponds to the $\Pub$ associated with the locking script, hence claim the output.

The inputs $\{ \utxo_1,\utxo_2,\cdots,\utxo_n \}$ represent unspent transaction outputs claimed by the buyer from previous transactions.
When a buyer decides to use a specific output from a previous transaction as an input to a new transaction, the buyer must specify proof that he/she still possesses ownership rights and did not previously spend them in another transaction.
This is done by attaching an \textit{unlocking script} to each input.
The unlocking script solves the locking script that was associated with the output from the previous transaction.
Likewise, the unlocking script is a digital signature produced by the $\Pr$ of the buyer that corresponds to a $\Pub$ associated with the locking script of an $\utxo$.
A valid unlocking script is legitimate proof of continuous possession of ownership rights to certain BTC being used as input.
As a result, BTC can be viewed as a chain of digitally signed transactions where ownership rights are transferred from one owner to the other by digitally signing them.

A transaction must include at least one input, however, may include multiple outputs to simultaneously pay different sellers from the total value associated with the inputs.
The locking script of each output would specify the conditions of its claimer.
However, it is necessary that the total BTC value of the inputs is always equal to or greater than the total value of the outputs.
In the event that the total value of the inputs is greater than the total outputs, the difference, known as the \textit{transaction fee}, is rewarded to the miner that adds the transaction into a block attached to the blockchain.
For guaranteed processing, most available wallets today derive the transaction fee as a fixed amount of BTC in relation to the size of the transaction.
In other words, the transaction fee increases with the size of the transaction.

The wallet of the user combines all the transaction inputs/outputs and their corresponding scripts into one digital message $\M$.
It then applies the Secure Hash Algorithm $\SHA$ to $\M$ twice to increase security before releasing it into the network.
The 32 byte digest representing the identity of the transaction ($\id_{\tx}$) is generated as follows
\begin{equation} \label{eq:id}
\id_{\tx} = \SHA\left(\SHA(\M)\right).
\end{equation}

A newly generated transaction assembled by the BTC wallet of a buyer is released into the Bitcoin network to be validated and stored in the blockchain.
The generating node transfers the transaction to its peers which flood it to the rest of the network nodes.
Each node that receives it audits the inputs by executing the scripts associated with it.
This audit involves checking whether the execution of the unlocking script integrated by a buyer within each input matches its corresponding locking script defined in the previous transaction. 
If a match exists, the node relays the transaction to its peers and temporarily places it in its \textit{transaction pool} until chosen to be mined, otherwise, the transaction is dropped.

In some cases, transactions are not flooded into the network in the same order they are generated.
As a result, during the audit, a node might not be aware of some inputs of a transaction (child transaction) referring to the outputs of other transactions (parent transactions).
Instead of immediately rejecting the transaction and considering its inputs as invalid, the node can temporarily place it into an \textit{orphan transaction pool}.
If the parent transaction shows up, the inputs of the child transaction become valid and it can be transferred to the transaction pool.

\subsection{Bitcoin Transaction Standards}   \label{subsec:transactiontypes}
Currently, there are five Bitcoin transaction standards and a few non-standard transactions.
All transaction types are generated with a stack-based scripting language that is processed from left to right.
A script consists of a list of instructions that must be executed in the correct order to grant an individual the right to spend the BTC within a transaction. The list of standards is described below.

\para{Pay to Public Key Hash (P2PKH)} 
This standard transaction is the most used type.
The locking script within each output of a transaction holds the public key hash (serving as a Bitcoin address) of the seller that will claim the BTC amount included.
In other words, the locking script defines a condition that the seller must possess a specific $\Pr$ corresponding to the public key hash to claim the output.
Once claimed by the seller, the output becomes an $\utxo$ owned by the seller.
In order for the seller to use this specific $\utxo$ as an input to a future transaction, the seller must attach a valid unlocking script to it.
The unlocking script includes the $\Pub$ of the seller and a digital signature generated by his/her $\Pr$ that corresponds to the public key hash associated with the locking script of the previous transaction output.

\para{Pay to Public Key} 
The intent behind this standard transaction is to simplify the P2PKH standard.
Rather than associating the public key hash within the locking script of the output, the public key itself is used.
As a result, the validation process is simple. 
The digital signature of the seller generated with a $\Pr$ can immediately be compared to the associated $\Pub$ by searching whether or not they match.

\para{Multi-signature (MultiSig)}
In this standard transaction, a combination of keys is required to authorize an output claim.
The locking script of a transaction output is associated with a number ($N$) of public keys.
In order for an individual to claim the output, the individual must possess $M$-of-$N$ private keys that correspond to the $N$ public keys.
This type of transaction can increase the security and can be used in scenarios which require more than one user to be present in order to claim and spend BTC.
However, as the number $N$ of public keys associated with the transaction output increases, the size of the transaction also increases.
As a result, these transactions acquire large space in the $\utxo$ pool, therefore requiring more storage memory.	
As discussed previously, larger transactions also require larger transaction fees.

\para{Pay to Script Hash (P2SH)}
This standard transaction was introduced to resolve the complex issues caused by MultiSig transactions.	
The transaction has the same simple complexity as a P2PKH transaction.
Rather than associating the entire locking script with a transaction output that includes multiple public keys, a double hash computation is applied to the entire script, specifically $\SHA\big(\RIPEMD(script)\big)$.
The result is a 20-byte digest that is attached to the locking script instead of the entire original script.
In order to use the output from this transaction as an input to another transaction, the buyer creates an unlocking script that holds $M$-of-$N$ private keys and the original script that was cryptographically hashed earlier.
In that way, sufficient information is available in the locking and unlocking scripts to validate the $\utxo$ for spending.
In addition, the buyer no longer has to worry about generating large transactions which might require hefty transaction fees to process.
Instead, only the seller is required to provide the unlocking script he/she wishes to spend the output in a new transaction.

\para{Data Output}
This standard transaction is intended to store arbitrary data on the blockchain rather than transfer BTC from a buyer to a seller.
In the Bitcoin community, many members believe that such transactions are abusive to the system since it places a burden on the network nodes to process transactions that do not carry BTC.
However, such transactions exist and allow 40 bytes of data to be stored per transaction.
These transactions are un-spendable, therefore are not stored in the $\utxo$ set.

\para{Non-Standard}
A very small percentage of transactions are processed under non-standard transactions.
Non-standard transactions use more sophisticated scripts that strive to provide higher complexity and security.
In some cases, these transactions might even be the result of bugs or mistakes resulting in loss of BTC.

\subsection{Merkle Trees}
Validated transactions are grouped into blocks which are then mined and stored in the blockchain.
A single block can contain multiple transactions up to the block size limit.
Merkle trees, sometimes referred to as hash trees, are utilized to cluster multiple transactions in one block.

A Merkle tree is a tree data structure generated in a bottom-up approach that can efficiently summarize and verify the integrity of the transactions being combined.
Starting from the leaf nodes which are hashes of the original data, each non-leaf node is generated as a computation of its respective children nodes.
For a single non-leaf node, all its children nodes are concatenated then hashed to produce a single digest that represents the node in the tree.
This approach continues until a single node is generated which is defined as the root node.

BTC utilizes a binary Merkle tree in which each non-leaf node has exactly two children.
It applies a double hash computation $\SHA\left(\SHA(\cdot)\right)$ when generating nodes.
The leaf nodes used to construct the tree are the identities $\id_{\tx}$ generated for each transaction as discussed in equation~\eqref{eq:id}.

In a binary Merkle tree, each row within the tree consists of an even number of nodes, except the root node.
In the case where a row consists of an odd number of nodes, a replica of the last node is reproduced to even out the number of nodes in that row.
To better comprehend the construction of the binary Merkle tree, consider a block that consists of five transactions, $\{ \tx_1,\tx_2,\cdots,\tx_5 \}$.
Each one of these transactions has already been validated by the nodes and an identity for each transaction has been generated as discussed in equation~\eqref{eq:id}.
We denote the corresponding identities as $\{ \id_{\tx_1},\id_{\tx_2},\cdots,\id_{\tx_5} \}$, where each identity represents a leaf node in the tree.
In this example, the number of nodes at the leaf node level is odd, therefore a replica of the fifth identity is generated, $\{ \id_{\tx_1},\id_{\tx_2},\cdots,\id_{\tx_5},\id_{\tx_6}=\id_{\tx_5} \}$.
Next, the double hash computation is applied to the concatenation of each two identities to generate the parent non-leaf nodes of the Merkle tree as follows
\begin{align} \label{eq:digests1}
N_1 &= \SHA\big(\SHA(\id_{\tx_1} \| \id_{\tx_2})\big),\\
N_2 &= \SHA\big(\SHA(\id_{\tx_3} \| \id_{\tx_4})\big),\\
N_3 &= \SHA\big(\SHA(\id_{\tx_5} \| \id_{\tx_6})\big),
\end{align}
where $\|$ is the concatenation of two identities.

As shown in the previous equations, an odd number of non-leaf nodes is generated at that level.
To even it out, we replicate $N_3$ to produce $N_4$ as

\begin{equation}  \label{eq:digests2}
N_4 = \SHA\big(\SHA(\id_{\tx_5} \| \id_{\tx_6})\big).
\end{equation}

Using the resulting digests we can generate the following level of non-leaf nodes as
\begin{align} \label{eq:digests3}
N_5 &= \SHA\big(\SHA(N_1 \| N_2)\big),\\
N_6 &= \SHA\big(\SHA(N_3 \| N_4)\big).
\end{align}

Finally, the 32 bytes root node is derived as
\begin{equation}  \label{eq:digests4}
R = \SHA\big(\SHA(N_5 \| N_6)\big).
\end{equation}

\begin{figure}[h]
\centering
\includegraphics[scale=0.45]{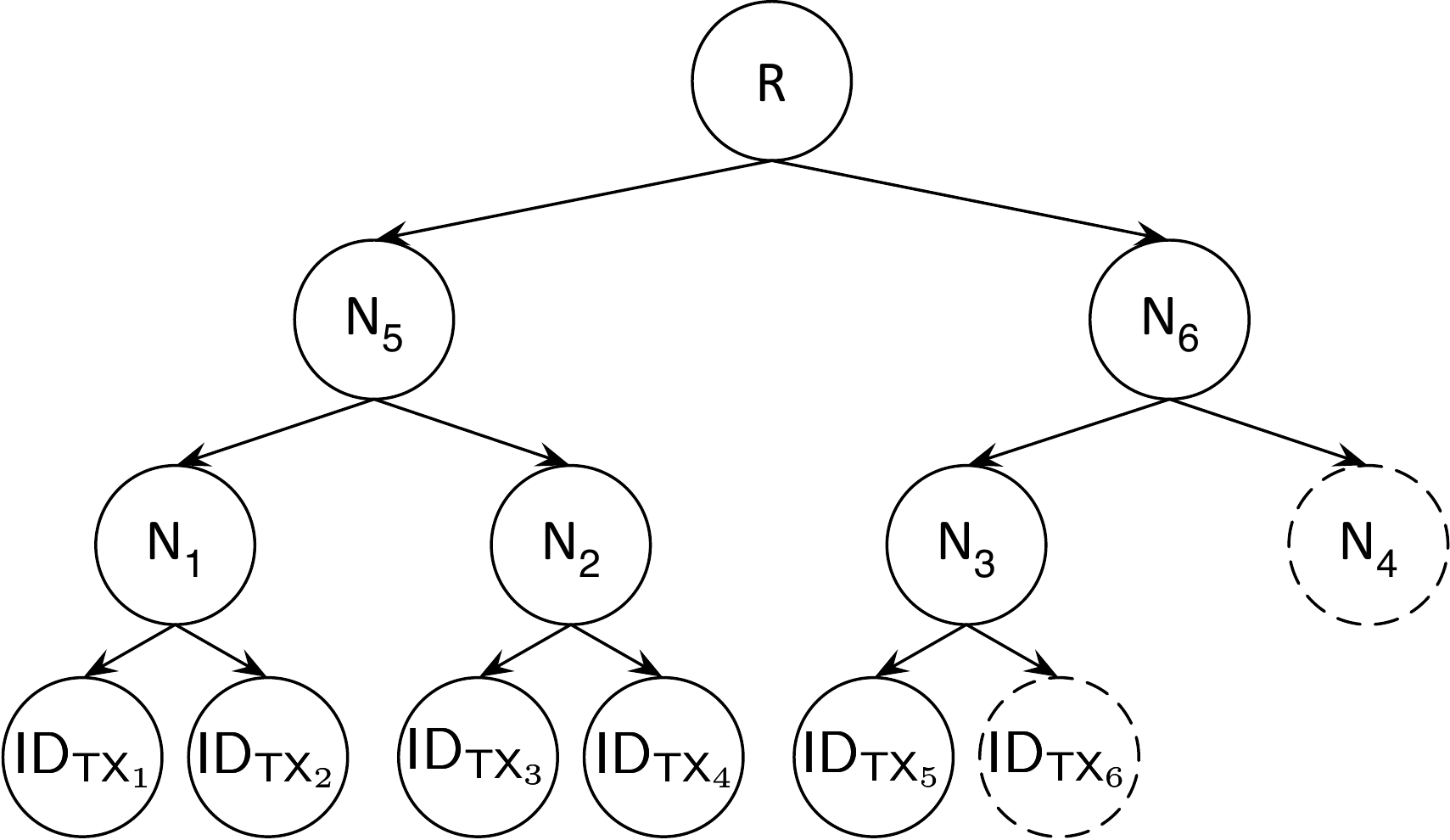}
\caption{A Merkle tree within a block.}
\label{Fig:merkle}
\end{figure}

Fig. \ref{Fig:merkle} represents the complete construction of the Merkle tree for this example. 
The dotted nodes represent the replicated nodes that are added to even out the odd rows. 
The root node, R, representing the summary of all transactions is placed into the block header of a block to be mined and chained to the blockchain. 

The use of Merkle trees is more common in SPV nodes since they do not store a copy of the full blockchain.
When an SPV node needs proof to the existence of a transaction within a block, it turns to a full node for assistance. 
The full node will generate a \textit{merkle path} by computing a maximum of $\log_2N$ $\SHA\big(\SHA(\cdot)\big)$ computations, where $N$ represents the total number of transactions in the tree.
Using the Merkle path as an authentication path, the SPV node can prove the existence of a transaction within the tree.
This proof of existence method is considered to be efficient since it only requires hash computations.

\subsection{Blockchain}

The blockchain is a public ledger that stores all previous transactions since the creation of Bitcoin.
It provides its users with transaction confirmations to track ownership rights of BTC.
As new transactions are processed, the blockchain is extended.
It consists of blocks $\{ \B_0,\B_1,\cdots,\B_n \}$, each carrying a set of validated transactions, where $\B_0$ represents the first block and $\B_n$ represents the most recent block attached to the blockchain. 
Blocks are linked back-to-back, with each one referencing its previous block to form the complete blockchain.
To reference a block, a unique 32 byte identity $\id_{\B_i}$ is generated for $\B_i$ by applying $\SHA\big(\SHA(\cdot)\big)$ to the block header.
An identity is referred to as the block hash.

The head of the blockchain is denoted as $\B_0$ and is defined as the \textit{genesis block}.
$\B_0$ differs from all the other blocks as it does not reference any previous block.
At the launching stage of the system, $\B_0$ was a stand-alone block waiting for the system to initiate a new mined block to be chained to it.

Each block consists of two parts, a header, and a body.
Each header incorporates the block hash of its predecessor block in the chain.  
The header also consists of a difficulty target, nonce, and a time-stamp which are discussed in more detail in the following subsection.
The body carries all the leaf nodes and non-leaf nodes of the Merkle tree, excluding the Merkle root, which is incorporated in the header.
This design makes it infeasible to retroactively alter records within any block of the blockchain.
Any modification to one block will require adjusting all the subsequent chained blocks.

\subsection{Bitcoin Mining} \label{subsec:bitcoin_mining}
Bitcoin mining is the final stage to secure validated transactions and add them to the blockchain.
Once a transaction is added to the blockchain, it becomes completely verified and public to all users.
The transaction claimer(s) can use the embedded $\utxo$(s) as the input to other transactions whenever desired.

Miners begin by selecting transactions from their transaction pools that will be placed into a block where a block cannot exceed 1MB in size. 
A small portion of that space is specified to carry high priority transactions. 
Priority is based on the size and age of the transaction inputs.
The rest of the block is filled with other transactions which have greater transaction fees to maximize the profit that a miner can turn if successful in mining the block first.
A transaction with low or no fees will probably remain in the transaction pool of the miner until it ages and becomes a high priority transaction.

Next, the miner assembles a special transaction, known as the \textit{coinbase transaction}.
This transaction is a reward paying transaction to the miner in the event of winning a mining competition.
It does not have any inputs and consists of a single output addressed to the wallet of the miner.
The amount incorporated in the output is the reward mining fee (12.5 BTC at the time of writing) plus the sum of all transaction fees included in each transaction.

All the selected transactions along with the coinbase transaction are then combined into a Merkle tree as discussed previously.
At this point, the miner has all the components needed to construct the block header of the new block except the \textit{nonce}.
The nonce is a value which if concatenated with the block header of the group of chosen transactions and then double hashed, it produces a digest with a specific prefix of zeros in its binary representation.
Searching for this value is performed in a brute-force manner and is directly correlated with the computational power available.
The more available computational power, the faster a miner is able to find the correct nonce.
A successful miner will then broadcast his/her \textit{proof-of-work} to prove that he/she consumed computational resources in order to find the correct nonce.

The primary advantage of the proof-of-work is to make it computationally infeasible to perform Sybil attacks~\cite{douceur2002sybil}.
This process is intentionally designed to be resource-intensive to perform while efficient to verify that the work has been done.
It is required that a certain number of zeros appear in the prefix of the digest as a result of applying the double $\SHA$ computation.
The prefix determines the \textit{difficulty} of finding the correct nonce.
The more zeros required in the prefix of the digest, the harder it is to find the correct nonce and vice versa.
The difficulty is dynamically altered every two weeks so that the average time it takes a miner to find a correct solution is approximately ten minutes.
As the number of miners increases, the difficulty increases, and vice versa.

The first miner to find the correct nonce to a block of transactions is rewarded a mining reward as compensation for the computational power spent.
The mining reward is halved precisely every 210,000 blocks that are added to the blockchain.
It is estimated to continue until the year 2140 when nearly 21 million BTC will have been released into the system.
The reason for having a fixed supply of BTC is to prevent price inflation in the future.

Another incentive that encourages miners to spend their computational power to perform mining is the transaction fee.
The winner is not only rewarded the mining reward but is also given all the transaction fees incorporated with all the transactions in the block.
With time, the mining reward will decrease due to halving, which will demand higher transaction fees in the future to compensate for the reduced mining reward.

After a block is successfully mined, all the miners check their transaction pools to eliminate the transactions that have been included in the mined block and immediately construct a new block of transactions.
The end of the mining race marks the beginning of a new one.
Miners instantly begin to search for the nonce of the next block of transactions.

Simultaneously, the mined block is flooded through the network so that all the nodes can update their blockchains.
The winning miner transmits the block to its peer nodes to validate it before propagating it further through the network.
The peer nodes check whether the block is correctly assembled in terms of syntax and variables.
The proof-of-work provided by the miner must be correct and the coinbase transaction must include the correct amount to pay the miner.
If any information is invalid, the block is immediately dropped.

Quite regularly, as blocks are mined to extend the blockchain, a temporary incident, known as a \textit{fork}, might occur.
A fork occurs when two miners are able to simultaneously mine two different blocks at the same time.
As a result, both newly mined blocks are accepted to extend the blockchain.
The blocks are flooded into the network and the miners update their version of the blockchain based on the block they receive first.
This results in two valid versions of the blockchain in possession by the miners with two different paths.
However, the miners continue to extend their version of the blockchain regardless which path they possess.
Eventually, one path will grow longer than the other as mining continues.
The path that grows longer is the winner and all nodes immediately discard the other path and update their blockchain to the longer one.
In literature, the blocks that are dropped are known as \textit{orphan blocks}; valid blocks that were part of the blockchain at some point.

\subsection{Bitcoin Mining Pools and Payment Methods} \label{subsec:payment}
Although solo miners can compete in the mining process, the likelihood of a successful return is very low.
This is even the case for solo miners with the most powerful computing machines.
As a solution to this problem, solo miners collaborate in the mining process by joining computational power into mining pools.
Together, they form a large organization with significant computational power that can compete with the other large entities.
The members of the mining pool work together to find the correct nonce for a candidate block and report the result as one miner, increasing their chances of winning the competition.
In the event of success, the rewards are split among the participating miners based on the contribution provided by each.

The concept of a mining pool can be compared to the lottery.
Assuming individuals with the same financial capabilities, if a large group buys tickets together, the individuals within the group have a better chance of winning than a single individual buying tickets alone.
If any ticket owned by the group wins the lottery, the participating individuals split the reward proportional to the amount invested by each.
In a mining pool, the computational power provided by each solo miner is analogous to the amount invested by each ticket buyer.

A mining pool is managed by a pool operator who handles the entire pool server and receives a percentage of the rewards as compensation.
The role of the operator is to coordinate the mining performed by all the participating miners.
The operator keeps a continuously updated copy of the entire blockchain to ease the job of the participating miners.
Using the updated blockchain, the operator verifies any transaction that appears in the network and places it in a candidate block for mining.
By that, miners only need to worry about finding the correct nonce of the candidate block.
If the mining pool wins the competition, the operator divides the rewards among the participating miners.

Reward splitting can be performed in multiple forms and varies from one mining pool to the other. 
As described in~\cite{rosenfeld2011analysis}, these methods can be categorized into simple reward, score-based reward, or risk-free pay-per share reward.

Simple reward systems consist of either proportional systems or Pay-Per Share (PPS) systems.
In the proportional systems, a reward $B$ is split among the participating miners at the end of each round, where a round is the consecutive time between two successful blocks generated by the pool.
The operator keeps a percentage of the reward $fB$ and divides the remaining $(1-f)B$ among the miners based on the shares they submit. 
Shares are defined as the number of hashes performed by each miner in attempt to find the correct proof-of-work. 
A miner that submits $n$ shares from a total of $N$ shares submitted by all the miners in the pool receives a reward of $\frac{n}{N}(1-f)B$ BTC on average.
Conversely, the PPS system is a deterministic one where the miner knows how much reward can be turned in advance.
The operator immediately pays each miner based on the submitted shares regardless of the mining result.
In other words, a miner that submits $n$ shares receives $(1-f)pB$ BTC/share, where $p$ represents the probability of one share being the correct proof-of-work.
In this system, the operator is taking the risk of mining independently since the miners receive ensured payments whether or not the pool generates a block.

Score-based reward systems come in many forms and strive to prevent miners from pool-hopping.
Pool-hopping is the practice of mining in a pool only during its good times (successfully generating blocks) and leaving it during its bad times.
A pool-hopper can maximize his/her rewards at the expense of miners that remain loyal to the pool at all times.
The method introduced by Slush~\cite{slush} is one of the first implemented score-based systems that extends the proportional method.
Rather than paying the miner an amount based on the submitted shares after each round, the miner is given a score that is proportional to his/her contribution and increases as more time elapses from the start of the round. 
The score is used to calculate the reward share given to the miner at the end of the round.
However, this method is still susceptible to hopping since the score does not consider factors such as the mining difficulty or the hashrate of the pool.
Also, in this method mining at the beginning of a round is more profitable since there are fewer shares at that time.
As a result, the geometric method was introduced to address these weaknesses.
This method introduced a fixed fee, a constant amount taken from the reward of each block, and a variable fee, a score granted at the beginning of each round to the operator.
As time passes, the variable fee declines, making mining equally profitable throughout the entire round.
Shorter rounds result in larger variable fees and vice versa.
By that, there is no advantage to mining early in the round.

Another score-based method is Pay-Per-Last-N-Shares (PPLNS) that exists in different forms.
In this method, the concept of rewarding miners after each round is replaced with rewarding miners that have been participating in earlier rounds, regardless of the mining result.
In other words, the operator pays miners based on their contributions from previous efforts.
Later on, more advanced payment systems evolved such as the Double Geometric Method (DGM).
This system is a hybrid between the PPLNS and geometric system that combines advantages of both methods.

Some mining pools employ a risk-free pay-per share system.
One of the first implemented systems is known as the Maximum Pay-Per Share (MPPS). 
It combines both the PPS and proportional systems, where each participating miner has a balance of each.
If the miner submits a share, the PPS balance is incremented and when the pool successfully generates a block, the proportional balance is incremented.
At pay time, the miner receives the minimum of both balances.
This method protects the pool from taking the risk alone.
However, this method is inconsiderate to the miners, since they will always make less whether the pool is successful or not.
In addition to this, the system suffers from pool-hopping.
A solution was later proposed to solve this problem in the Shared Maximum Pay-Per-Share (SMPPS) system.
The miners have a PPS balance which continues to accumulate as the miners are participating.
If a block is found by the pool and there are sufficient funds, the miners are paid based on their PPS balance.
However, if there are no sufficient funds, miners are paid proportional to the available funds and given credit to be paid later for whatever balance that is owed.

Today, a broad range of mining pools exist that give miners a variety of options when joining pools.
The question most miners would ask is which mining pool is the best to join.
The answer here lies in the preferences of the miners.
For example, some miners are not willing to take the risk of not getting paid in the event of being unsuccessful in generating a block and would prefer a PPS mining pool.
Others might be willing to take the risk and choose a score-based system for instance, in return for larger profit.

\subsection{Alternative Cryptocurrencies}

In literature, alternative cryptocurrencies are known as \textit{altcoins}, most of which are inspired by Bitcoin.
Altcoins strive to offer innovative features and/or enhanced security/privacy countermeasures in an effort to compete with Bitcoin.
Their development process is based on the level of innovation and security/privacy countermeasures they present.

The simplest method to develop an altcoin is by forking the open source code of Bitcoin~\cite{bitcoincode} while adding/modifying any features to it.
In software development, a \textit{fork} is a completely independent project that exploits a copy of the original source code. 
A Bitcoin fork generates an entirely new blockchain and is completely independent of Bitcoin.
Namecoin~\cite{loibl2014email} is the first developed Bitcoin fork that adopted all of the characteristics of Bitcoin. 
It also introduced an additional feature allowing users to store data within its transactions.
Various Bitcoin forks have evolved latterly with more features and handled security/privacy issues.
Many of these forks implemented privacy protocols to increase the anonymity of cryptocurrencies.
In Section~\ref{Sec:privacy}, we discuss notable privacy protocols that have impacted some of these altcoins.

In exceptional occasions, an altcoin can also be the result of a \textit{hardfork}.
A hardfork occurs when modifications are made to the original software of Bitcoin making its new transactions/blocks incompatible with those previously generated prior to the modifications.
These modifications can be as simple as altering certain parameters, such as the block size, or as complex as changing major protocols, such as the consensus algorithm.
In order to enforce these modifications, the majority of users/miners must upgrade their client nodes to the latest version which accommodates these changes.
The users/miners that do not accept the upgrade will view the new transactions/blocks as invalid and will not accept them.
As a result, the blockchain will inevitably split into two paths, one storing transactions of the original cryptocurrency and one storing transactions generated due to the modifications made, hence creating a new altcoin.
Users in possession of the original cryptocurrency will automatically be granted an equivalent amount of the new altcoin to what they hold.

Bitcoin Cash is a notable example of a Bitcoin hardfork which occurred on August 1, 2017.
It was the result of enforcing $\bip91$~\cite{bip91} which proposed activating Segregated Witness (SegWit)~\cite{wuille2015segregated}.
SegWit increases the transaction speed of Bitcoin by splitting the transaction into segments and removing the unlocking signatures which are attached separately at the end. 
The majority of the miners accepted this proposal resulting in Bitcoin Cash.
Users who possessed BTC were immediately granted an equivalent amount of BCC (The currency of Bitcoin Cash) to the BTC they possessed.

While only borrowing the concept of storing transactions in a blockchain, some altcoins have been implemented from scratch with a completely different design and purpose.
These altcoins strive to provide services and security/privacy countermeasures beyond the capabilities of Bitcoin or any of its forks.
They present substantial differences such as integrating enhanced consensus algorithms or utilizing private (permissioned) blockchains.
In contrast to the public (permissionless) blockchain of Bitcoin, where all participating nodes are allowed to execute the consensus protocol and maintain the blockchain, a private blockchain is limited to only specific nodes.
As a result, the cryptocurrency market has witnessed a considerable number of altcoins with substantial innovative features.

\subsection{Major Security and Privacy Issues}

Cryptocurrencies are regarded as robust transacting systems designed to avoid payment fraud and provide superior user privacy.
However, major issues in cryptocurrencies have been theorized that can put security and privacy in jeopardy.

First, the attacker may potentially deceive the system by spending the same coin more than once.
This is known as the double spending attack.
Second, cryptocurrencies may be exposed to further weaknesses through major network and storage vulnerabilities. 
Third, while cryptocurrencies strive to provide their users with anonymity, the current solutions could be vulnerable to linkage problems putting the privacy of the users in jeopardy. 

In the following sections, we aim at highlighting major security and privacy issues that exist in Bitcoin and similar cryptocurrencies today.

\section{Double-Spending Attacks} \label{Sec:DS}
Double-spending is an attack that could be performed by malicious users attempting to deceive the system by spending the same BTC more than once.
The attacker generates duplicates of the same $\utxo$ and uses it as an input in more than one transaction.
Differentiating between the duplicated (fraudulent) copies and the original becomes an issue when used in a decentralized system. 
There is no trusted entity that verifies the legitimacy of the $\utxo$ used as input in a transaction.
The inputs of a transaction may consist of unidentifiable fraudulent BTC that have possibly been spent earlier.

The system defends against such attacks by relying on its users (miners) to validate the legitimacy of the BTC used as an input to transactions.
Using the information stored in the blockchain from the previous transactions, the miners validate the inputs of any new transaction to ensure that it does not contain previously spent inputs.
Once verified, the transaction is mined into a block which is attached to the blockchain.
Any user that refers to the blockchain becomes aware that specific $\utxo$(s) have been spent earlier, making fraudulent input transactions detectable.

To ensure that attackers cannot manipulate the blockchain in their favor, the mining process is designed to be an expensive and resource-intensive operation.
To mine a block of transactions in the blockchain, the miners must provide a valid proof-of-work.
An attacker that wishes to double-spend BTC must reverse a transaction that has been stored in the blockchain to reuse its inputs in another transaction.
Reversing an already stored transaction in the blockchain is an extremely difficult task since it requires a significant share of the total computational power of the system.

In the rest of this section, we will analyze the double-spending attacks.
We first discuss conventional methods to perform double-spending. 
Next, we analyze the probability and profitability of the double-spending attack and present a trade-off between the waiting time before accepting a transaction versus the profitability of the attack.

\subsection{Types of Attacks}
A double-spending attack comes in many forms. We discuss various techniques that can be performed.

\subsubsection{Race Attack}
A race attack refers to the case where a merchant accepts an unconfirmed transaction (a transaction in a transaction pool waiting to be mined and stored in the blockchain) and immediately provides the payer with a product/service before waiting for confirmation.
An attacker with the intention of deceiving the merchant creates two transactions: (i) a transaction that pays the merchant an amount of BTC in return for a product/service and (ii) a fraudulent transaction that pays the same amount to the wallet of the attacker.
Both transactions use the same inputs (duplicated BTC) and try to spend the same BTC.
The attacker concurrently releases both transactions into the Bitcoin network.
The miners consider both transactions as being valid until one of them gets stored in the blockchain.
The transaction that gets stored in the blockchain is referred to as a confirmed transaction.
At that point, the inputs of the stored transaction cannot be used as inputs to other transactions.
Therefore, the fraudulent transaction has a chance of being verified first and added to the blockchain making the merchant-paying transaction invalid.
The invalid transaction is rejected by the system and dropped from the transaction pools of miners.

To avoid a race attack, merchants must wait for the mining to be completed and the transaction to appear in the blockchain before providing the payer with the product/service. 
It is recommended that the merchant should wait for at least six subsequent blocks as confirmation before making the trade.
In this case, the chances for an attacker to reverse a transaction are negligible, assuming that the attacker can control no more than 10\% of the total computational power used in mining.

\subsubsection{Finney Attack}
Finney attack was first suggested in a Bitcoin forum~\cite{finney}. 
Similar to the race attack, the attacker performing this attack will only succeed if the merchant accepts an unconfirmed transaction. 
The attacker creates two transactions similar to those in the race attack and holds on to both of them.
The attacker then begins mining the block containing the fraudulent transaction.
If the attacker is successful in mining the block, the attacker then uses the other transaction to pay a merchant immediately in exchange for a product/service.
Once the merchant makes the trade, the attacker releases the mined block which contains the fraudulent transaction into the network.
Given that the block is already mined, it will be added to the blockchain immediately.
As a result, the merchant-paying transaction will become invalid.  
In addition to this, the attacker is rewarded the mining reward for the mined block carrying the fraudulent transaction.
However, the ability to independently mine a block is improbable given the resources necessary to perform the task.

\subsubsection{Vector76 Attack}
In comparison to the race and Finney attacks, the Vector76 attack requires the merchant to wait for a single block to be mined and added to the blockchain as a confirmation.
To reverse the transaction, the attacker needs to create a fork in the blockchain.
Initially, the attacker creates a merchant-paying transaction and does not broadcast it to the network.
Next, the attacker tries to independently and secretly mine this transaction into a block.
If successful, the attacker holds onto the block until the honest miners discover another block.
The attacker then simultaneously releases the block into the network at the same time as the honest miners release their block which will result in a fork.
Before the fork is resolved, the attacker creates a fraudulent transaction that double-spends the same BTC used in the merchant-paying transaction.
The attacker then relays the fraudulent transaction to the honest miners that do not have the path of the blockchain that carries the merchant-paying transaction.
These miners see the fraudulent transaction as valid and begin mining it into a block.
As a result, each path of the blockchain stores one of the transactions.
If the path that holds the fraudulent transaction grows longer than the other path, the double-spending attempt is successful.

\subsubsection{51\% Attack}
51\% attack is the largest threat to the BTC system.
This attack is also referred to as the majority attack in which the attacker (usually a pool of miners) controls more than half of the total computational power of the system.
By controlling the majority of the power, the attacker is capable of interfering with the process of mining blocks and reversing any block of transactions.
During a 51\% attack, the system loses integrity since the other miners no longer have an incentive to compete in the mining process.

To better comprehend this attack, consider the case where the attacker generates a merchant-paying transaction and releases it into the network.
The merchant waits for an appropriate number of confirmations before accepting the payment and making the trade.
Simultaneously, the attacker secretly begins to mine a block that contains a fraudulent transaction followed by more blocks to extend it.
Since the computational power of the attacker is more than the rest of the computational power of all the miners combined, the attacker can mine blocks in less time.
Once the merchant accepts the transaction, the attacker releases the secretly mined blocks to create a fork in the blockchain.
If the fraudulent fork created by the attacker is longer than the original chain, it becomes dominant and all miners begin to extend on it.
By that, the merchant-paying transaction no longer exists in the blockchain.

This attack represents the biggest threat to Bitcoin as it is directly correlated to the resources an attacker can provide.
Resources are measured in terms of financial and computational power.
Large entities such as governments or intelligence agencies have the means to control a large share of the total computational power.
They are able to destroy or push the system into their favorable status.
It is important to note that even with a computational power that is slightly less than 50\%, an attacker may still be able to severely manipulate the system.
In the next subsection, we analyze the chances of success of the attackers based on the share of computational power they control.

\subsection{Probability of Success}
Despite the continuous increasing popularity of Bitcoin, the number of merchants that have accepted it as a method of payment today is still relatively minimal.
Many merchants have concerns about its capabilities in terms of security, while others consider it as a slow method to make payments. 
Those that accept it should try to take all precautions before accepting a transaction to prevent double-spending attacks.

One of the important precautions is to decide when to accept a transaction before making the trade.
Merchants prefer to obtain a certain degree of confidence as assurance that the payer will not be able to reverse the transaction.
Those that can afford to wait a long period of time before accepting a transaction (for example, online platforms) require a minimum of six confirmations before accepting a transaction and considering it as being irreversible.
However, others that cannot afford this time waiting (such as vending machines), rush into accepting transactions at the risk of losing the payment to a double-spending attack.

Similar to the analysis in~\cite{Nakamoto08Bitcoin}, we model the race between the honest miners and the attacker to generate blocks as a binomial random walk.
The race is denoted as $z$ which represents the number of blocks generated by the honest miners with computational power $p$ minus the number of blocks generated by the attacker with computational power $q=1-p$. 
If a block is generated by the honest miners, we increment $z$ by 1. 
Conversely, if a block is generated by the attacker, we decrement $z$ by 1.
The race between the honest chain and the chain generated by the attacker can be derived as
\begin{equation} \label{eq:random_walk}
z_{i+1} = 
\begin{cases}
z_i + 1, & \text{with probability } p,\\
z_i - 1, & \text{with probability } q,
\end{cases}
\end{equation} 
where $i$ represents an individual block race. 
If $q>p$ and the attacker has unlimited resources, the attacker will eventually reach $z<0$.
At that point, the attacker can replace the blocks generated by the honest miners and succeed in performing the attack.

The probability of the attackers to catch up and surpass the blocks generated by the honest miners can be compared to the Gambler's Ruin Problem. 
Similar to the description in~\cite{gamblers}, we assume a gambler (attacker) begins with an initial fortune $i$, $0<i<N$, and either wins \$1 with probability $q$ or loses \$1 with probability $p=1-q$, in each successive gamble.
The game represents a random walk which terminates at $i=0$ (fail) or at $i=N$ (success).
The probability of success after $i$ trials is denoted as $P_i$ and can be calculated as:
\begin{equation} \label{eq:gambler1}
P_i = qP_{i+1} + pP_{i-1}.
\end{equation} 

Since $q+p=1$, we can rewrite equation~\eqref{eq:gambler1} as:
\begin{equation} \label{eq:gambler2}
P_{i+1} - P_i = \frac{p}{q}\left(P_i - P_{i-1}\right).
\end{equation}

At $i=0$, the attacker has a probability of success $P_0=0$.
By rearranging and generalizing equation~(\ref{eq:gambler2}), we have
\begin{eqnarray} 
P_{i+1} &=& P_1 \sum_{j=0}^{i}\left( \frac{p}{q} \right)^j\nonumber\\
&=&  
\begin{cases}
P_1 \frac{1 - (\frac{p}{q})^{i+1}}{1 - (\frac{p}{q})} , & \text{if }  p \neq q,\label{eq:gambler3}\\
P_1(i+1) , & \text{if } p=q=0.5.
\end{cases}
\end{eqnarray}

Let $i=N-1$ meaning that $P_N=1$, we can rewrite equation~(\ref{eq:gambler3}) as
\begin{equation} \label{eq:gambler5}
1 = P_N = 
\begin{cases}
P_1 \frac{1 - (\frac{p}{q})^N}{1 - (\frac{p}{q})}, & \text{if }  p \neq q,\\
P_1N, & \text{if } p=q=0.5.
\end{cases}
\end{equation}

Solve $P_1$ from equation~(\ref{eq:gambler5}) and substitute the result into equation~(\ref{eq:gambler3}) to obtain
\begin{equation} \label{eq:gambler6}
P_i = 
\begin{cases}
\frac{1 - (\frac{p}{q})^i}{1 - (\frac{p}{q})^N}, & \text{if }  p \neq q,\\
\frac{i+1}{N}, & \text{if } p=q=0.5.
\end{cases}
\end{equation}

Following the analysis in~\cite{ozisik2017explanation}, we assume that the attacker begins with an initial fortune $i=y$ and can afford to lose up to $y$ dollars before giving up. The gambler wins if $i=N=y+z+1$ dollars. 
This assumption modifies the game to account for the probability $P_s$ of the attacker to surpass the blocks generated by the honest miners as
\begin{equation} \label{eq:gambler7}
P_i = 
\begin{cases}
\frac{1 - (\frac{p}{q})^y}{1 - (\frac{p}{q})^{y+z+1}}, & \text{if }  p \neq q,\\
\frac{y+1}{y+z+1}, & \text{if } p=q=0.5.
\end{cases}
\end{equation}

Consider an attacker that possesses an unlimited amount of resources and is willing to use as much of it as needed to perform the attack, i.e. $y \rightarrow \infty$.
If $q>p$, then
\begin{equation} \label{eq:gambler8}
\lim_{y\to\infty} \frac{1 - (\frac{p}{q})^y}{1 - (\frac{p}{q})^{y+z+1}} = 1.
\end{equation}

For $q<p$, we first divide the numerator and denominator by $\left(\frac{p}{q}\right)^y$ then calculate the limit as
\begin{equation} \label{eq:gambler9}
\lim_{y\to\infty} \frac{(\frac{p}{q})^{-y} - 1}{(\frac{p}{q})^{-y} - (\frac{p}{q})^{z+1}} = \left(\frac{q}{p}\right)^{z+1}.
\end{equation}

Finally, we can summarize the probability of the attacker to surpass the blocks generated by the honest miners as
\begin{equation} \label{eq:gambler10}
Q_z = 
\begin{cases}
\left(\frac{q}{p}\right)^{z+1}, & \text{if } q < p \text{ or } z \geq 0,\\
1, & \text{if } q > p \text{ or } z < 0.
\end{cases}
\end{equation}

The merchant has no way of figuring out the number of blocks that the attacker has been able to secretly mine.
Therefore, one way to model the overall probability of the attacker to surpass the honest chain is by using the Poisson distribution.
The expected number of blocks an attacker can generate is $\lambda = (z+1) \left(\frac{q}{p}\right)$.
The overall probability $P_s$ of the attacker to surpass the honest chain can be computed by multiplying the Poisson density and the probability of surpassing the honest $z-k$ remaining blocks as discussed in equation
\begin{align} \label{eq:success1}
P_s &= \sum_{k=0}^{\infty} \frac{\lambda^{k} e^{-\lambda}}{k!} \times Q_{z-k} \nonumber \\
&= 1 - \sum_{k=0}^{\infty} \frac{\lambda^{k} e^{-\lambda}}{k!}\!\times\!
\begin{cases}
1 - \left(\frac{q}{p}\right)^{z-k+1}, & \text{if } q < p \text{ or } k \leq z,  \\ 
1 - 1 =0, &\text{if } q > p \text{ or } k > z.
\end{cases}
\end{align}

For equation~(\ref{eq:success1}), if $q>p$, we will always have $P_s=1$, meaning that the attacker will win.
When $q<p$, the probability for the attacker to succeed is
\begin{equation} 
P_s = 1 - \sum_{k=0}^{z+1} \frac{\lambda^{k} e^{-\lambda}}{k!} \times \left(1 - \left(\frac{q}{p}\right)^{z-k+1}\right).
\end{equation}

Another way to model this probability is by using the negative binomial distribution assuming the attacker can pre-mine one block before broadcasting the merchant-paying transaction to the network~\cite{rosenfeld2014analysis}.
The merchant waits for $n$ blocks to be generated by the honest miners with computational power $p$ before accepting the transaction.
At that time, the attacker can secretly generate $m$ blocks with computational power $q=1-p$, where $m=n-z-1$. 
By definition, we can model this as the $m$ number of blocks that the attacker can generate (success) before the $n$ number of blocks the honest miners can generate (failure).
Therefore, the probability of a successful double-spending attack for a given value $m$ can be calculated as
\begin{equation}
P(m) = \binom{m+n-1}{m}\times p^nq^m.
\end{equation}

Overall, the probability for an attacker to successfully surpass the number of blocks generated by the honest miners can be computed as
\iffinal
\begin{align} \label{eq:sucess2}
\begin{split}
P_s &= \sum_{m=0}^{\infty}P(m) \times Q_{n-m-1} \\
&= 1 - \sum_{m=0}^{\infty}\binom{m+n-1}{m}\times p^nq^m \\
& ~~\times \begin{cases} 
1 - \left(\frac{q}{p}\right)^{n-m}, & \text{if } q < p \text{ or } k \leq n-m, \\
1 - 1 = 0, & \text{if } q > p \text{ or } k > n-m.
\end{cases}
\end{split}
\end{align}
\else
\begin{align} \label{eq:sucess2}
P_s &= \sum_{m=0}^{\infty}P(m) \times Q_{n-m-1} \nonumber\\
&= 1 - \sum_{m=0}^{\infty}\binom{m+n-1}{m}\times p^nq^m \times \begin{cases} 
1 - \left(\frac{q}{p}\right)^{n-m}, & \text{if } q < p \text{ or } k \leq n-m, \\
1 - 1 = 0, & \text{if } q > p \text{ or } k > n-m. 
\end{cases}
\end{align}
\fi

Similar to the previous analysis, equation~(\ref{eq:sucess2}) confirms that when $q>p$, the attacker will always succeed since $P_s=1$.
When $q<p$, the probability of success can be defined as
\begin{align} \label{eq:success_graphs}
P_s 
&= 1 - \sum_{m=0}^{n-1}\binom{m+n-1}{k}\times p^nq^m \times \left(1 - \left(\frac{q}{p}\right)^{n-m}\right) \nonumber\\
&= 1 - \sum_{m=0}^{n-1}\binom{m+n-1}{m}\times (p^nq^m - p^mq^n).
\end{align}

\begin{figure}[t]
\centering
\includegraphics[scale=0.51]{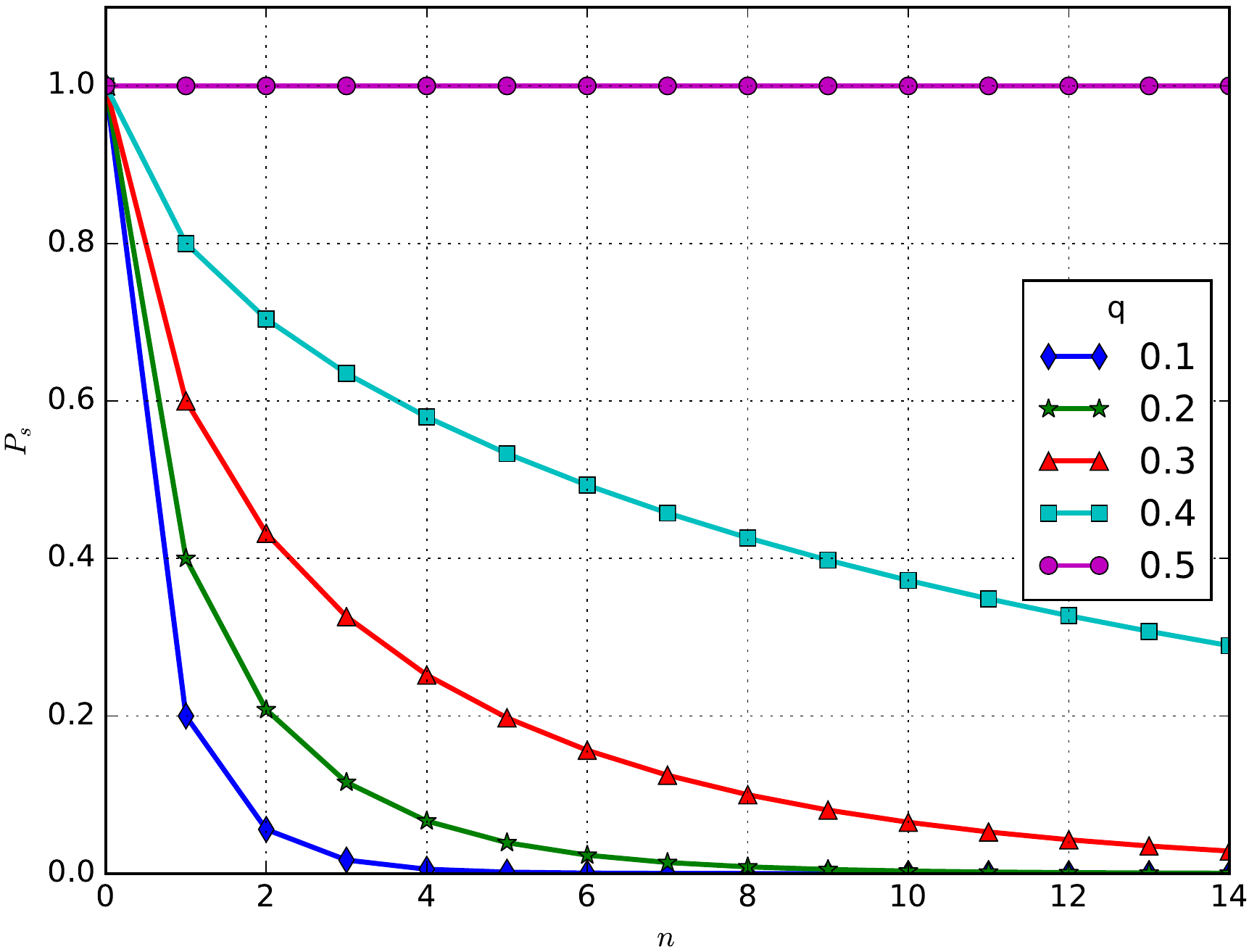}
\caption{Probability of successful double-spending attacks vs. number of confirmations waited by the merchant.}
\label{Fig:confidence_graph}
\end{figure}

Fig.~\ref{Fig:confidence_graph} shows the results of $P_s$ as $n$ changes based on equation~(\ref{eq:success_graphs}).
From this figure, the merchant can obtain the desired level of confidence before accepting a transaction. 
The obtained level of confidence is definite for any $q$ at $n=0$, meaning that the attackers have 100\% chance of success.
As the number of blocks $n$ increases, the chances of a successful double-spending attack decline. 
Conversely, as $q$ increases, the chances of a successful attack increase. 
The figure also shows that if $q\geq 0.5$ then we will always get $P_s=1$. 
This is known as the majority attack. 
In fact, even if the values of $q$ are slightly less than 0.5, the chances of a successful double-spending attack could still be high. 
However, the probability declines exponentially as the value of $n$ increases.

\subsection{Attack Profitability}
A successful double-spending attack is only profitable if the revenue is higher than the cost of performing the attack. 
Suppose an attacker tries to double-spend $v$ BTC paid to a merchant in exchange for a product/service. 
The attacker releases a transaction into the network that pays $v$ BTC to the wallet possessed by the merchant.
Immediately after releasing the transaction, the attacker secretly begins to mine blocks of transactions. 
One of these blocks contains a fraudulent transaction that pays the same $v$ BTC to the wallet possessed by the attacker. 
The merchant accepts the transaction after observing that $n$ blocks have been extended to the blockchain. 
If the attacker is able to secretly mine $m=n+1$ blocks and replace the $n$ blocks in the blockchain generated by the honest miners, then the attacker is successful in gaining a product/service without paying for it. 
Assume that the attack returns a value of $2v$, one of which is the actual BTC as a result of reversing the merchant-paying transaction and the other as the product/service.
In addition to this, the attacker gains the mining reward for each block mined and the transaction fees included in each transaction.
Then the revenue gained by the attacker can be formulated based on his/her corresponding $P_s$ as follows
\begin{equation} \label{eq:revenue}
\mathsf{Revenue} \approx v + P_s(v + Rm) ~\text{BTC},
\end{equation}
where $R$ is the block reward and the transaction fee per block. 

Multiple factors can impact the cost such as the price and depreciation value of machinery used, the cost of electricity, and the amount of BTC being spent in the transaction.
However, formulating the cost with all the possible factors is infeasible.
To simplify it, we focus our analysis on the cost factors that could change significantly as the attack is performed. 
These factors include the $v$ BTC an attacker spends in the merchant-paying transaction, the cost of mining $m$ blocks, and the depreciation cost $d(t)$ of the computing device used in BTC at time $t$.
We derive the cost as follows
\begin{align} \label{eq:cost0}
\mathsf{Cost}\approx v + me_q(t) + d(t)~\text{BTC} 
\end{align}
where $e_q(t)$ is the estimated mining electrical cost in BTC/block of a miner with a share $q$ of the total computational power of the system.
We assume $e_q(t)$ remains constant during the total time $T$ the attack is performed.

We also assume that the average lifespan of the mining equipment is approximately two years.
Using straight-line depreciation, $d(t)$ is a negligible value for an attack over a short period of time.
Therefore, we can reduce the cost equation as follows
\begin{equation} \label{eq:cost1}
\mathsf{Cost} \approx v + me_q(t)~\text{BTC}
\end{equation}

The time to mine a single block either by the honest miners or the attackers is approximately ten minutes.
Therefore, we can rewrite equations (\ref{eq:success_graphs}), (\ref{eq:revenue}), and (\ref{eq:cost1}) as
\begin{equation} \label{eq:success3}
P_s \approx 1 - \sum_{m=0}^{\frac{T}{10}-1}\binom{m+\frac{T}{10}-1}{m}\times (p^{\frac{T}{10}}q^m - p^mq^{\frac{T}{10}}),
\end{equation}

\begin{equation} \label{eq:revenue2}
\mathsf{Revenue} \approx v + P_s\left(v + \frac{RT}{10}\right) ~\text{BTC},
\end{equation}

\begin{align} \label{eq:cost2}
\mathsf{Cost} &\approx v + \left(\frac{T}{10}\right)e_q(t)~\text{BTC}.
\end{align}

The profit/loss can be formulated from equations (\ref{eq:revenue2}) and (\ref{eq:cost2}) as
\begin{align} \label{eq:profit}
\mathsf{Profit}/\mathsf{Loss} 
&= \mathsf{Revenue} - \mathsf{Cost} \nonumber \\
&\approx P_s\left(v + \frac{RT}{10}\right) - \left(\frac{T}{10}\right)e_q(t)  ~\text{BTC}.
\end{align}

Nowadays, to stand a chance in mining Bitcoin, miners merge their computational power into mining pools as discussed in Section \ref{subsec:payment}.
The mining pools combine the computational power provided by the computing machine of each participating miner.
Machines are categorized into one of four groups: Application-Specific Integrated Circuits (ASIC), Field-Programmable Gate Array (FPGA), Graphics Processing Unit (GPU), and Central Processing Unit (CPU).
Each group can provide up to a certain computational power.
Comparisons of most of those machines are presented in~\cite{hardware1} and~\cite{hardware2}.

Each computing machine consumes electricity differently based on its specifications.
Even machines with similar specifications might vary in cost to operate. 
As a result, formulating the cost of electricity spent by a miner in the mining process becomes challenging.

ASICs have monopolized the mining process due to their incomparable computational power with those of the CPUs, GPUs, and FPGAs.
Miners using any computing machines other than ASICs have a negligible chance of competing.
Many mining pools do not permit miners with these machines to join their pools.
A miner that joins a mining pool with one of these machines would hardly earn BTC in the event that the pool successfully mines a block. 
This is because rewards are usually divided among the miners based on their contributions as discussed in Section \ref{subsec:payment}.

Our goal now is to formulate the estimated electrical cost $e_q(t)$ of a mining pool.
First we estimate the total number of miners $N(t)$ based on the total hashrate $H(t)$ of the system at a certain time $t$ as
\begin{equation} \label{eq:numberofminers}
N(t) \approx \frac{H(t)}{h(t)},
\end{equation}
where $h(t)$ is the average hashrate of a single mining machine involved in mining at time $t$.

The cost of electricity is measured in cents/kWh and varies based on the end-use sector and time $t$. 
End-use sectors include, residential, commercial, industrial, and transportation.
We denote the average cost of electricity of all sectors at time $t$ as $e_a(t)$.
Using the computing wattage $w$ of the machine, the average running cost $c(t)$ of a machine at time $t$ is
\begin{equation}  \label{eq:runningcost}
c(t) \approx e_a(t) \times w ~\text{cents/hour}.
\end{equation}

Using equations (\ref{eq:numberofminers}) and (\ref{eq:runningcost}), the total cost $E(t)$ for all miners at time $t$ is
\begin{equation}  \label{eq:totalcost}
E(t) \approx N(t) \times c(t) ~\text{cents/hour}.  
\end{equation}

We know that it takes approximately 10 minutes to generate one block, i.e. in $T=1$ hour, miners can generate $m=6$ blocks.
Therefore, the total cost $C(t)$ for all miners to generate one block at $T = 10$ minutes can be estimated as
\begin{equation}  \label{eq:totalcostperblock}
C(t) \approx \frac{E(t)}{6}  ~\text{cents/10 minutes (1 block)}.
\end{equation}

However, as discussed previously, miners merge their computational power to increase their chances of winning in the mining competition.
The largest mining pool that exists today is Antpool~\cite{antpool} controlling approximately 25\% of the total computational power.
Other mining pools also exist such as BTCC Pool~\cite{BTTC}, Bixin~\cite{bixin}, BTC.com~\cite{btc}, and BTC.TOP~\cite{btctop} that control approximately 7\%-11\% of the computational power.

We estimate the average electricity cost $e_q(t)$ of a mining pool based on its computational power $q$ as
\begin{align} \label{eq:poolcost}
e_q(t) 
&\approx C(t) \times q \nonumber \\ 
&\approx \frac{H(t)\times e_a(t) \times w \times q}{6h(t)} ~\text{cents/block}.
\end{align}

For our simulations, we assume that the total cost of mining blocks $C(t)$ by all miners and computational power $q$ of the mining pool are fixed during the total mining time $T$.
We also assume a mining environment consists of miners using only ASICs such as Antminer S9 since it is one of the most efficient computing machines on the market today.
The specifications of this machine are $h = 14$ TH/s and $w=1.375$ kWh.

\begin{figure}[t]
\centering
\includegraphics[scale=0.5]{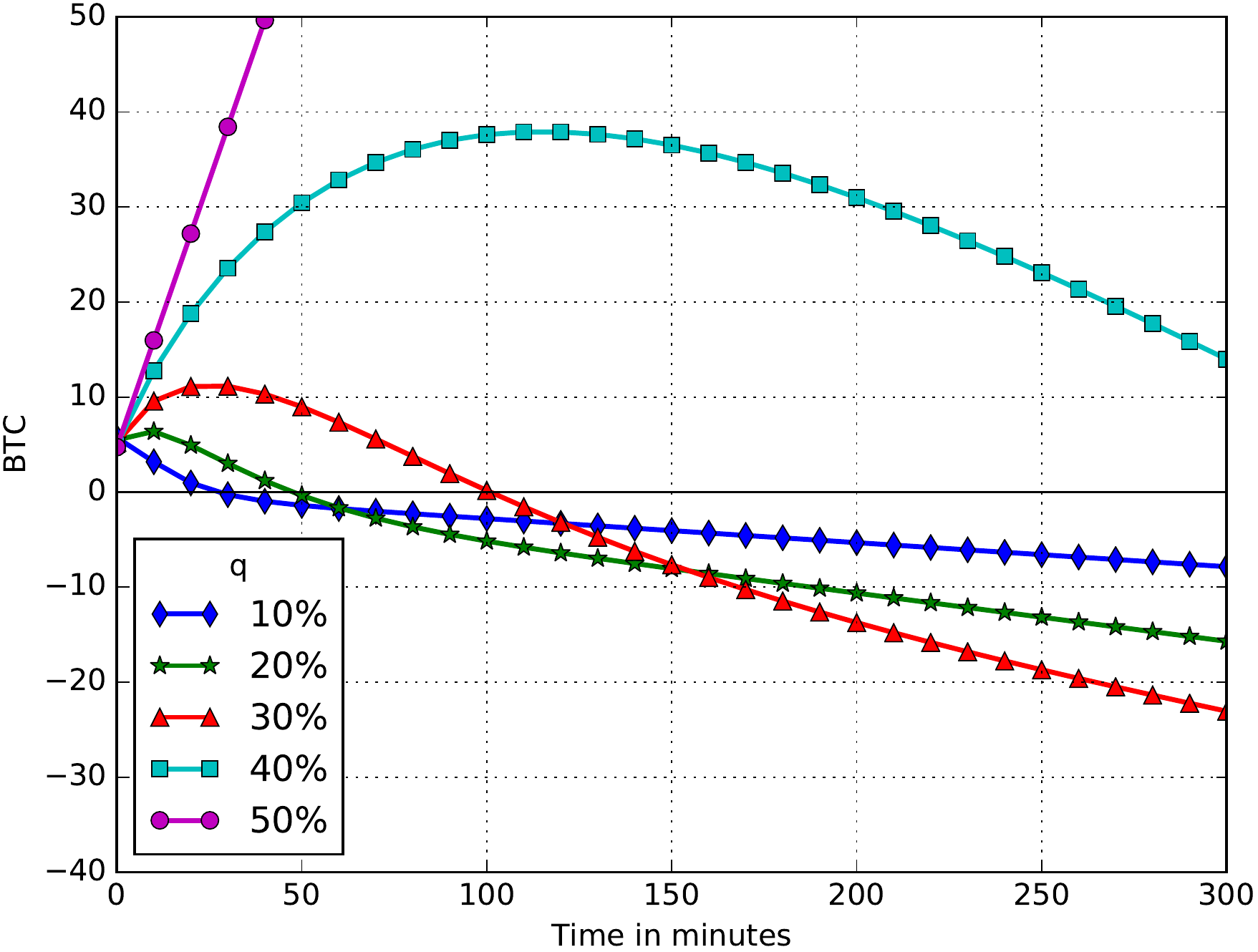}
\caption{Profit/loss of attackers with varying computational power $q$ trying to double-spend $v=5$ BTC.}
\label{Fig:profit}
\end{figure}

Consider an attacker trying to perform a double-spending attack during the period of August 2017.
During that period, 1 BTC was equal to approximately \$4500.
The total hashrate power was approximately $H=6,336,174$ TH/s and the average cost of electricity for all sectors in the U.S. was approximately 10.98 cents/kWh, based on the data collected by the U.S Energy Information Administration~\cite{eia}.
Under these circumstances, in Fig~\ref{Fig:profit} we present the expected profit/loss of double-spending attacks for various computational powers $q$. For this analysis, we assume the attackers try to double-spend $v=5$ BTC.

In Fig.~\ref{Fig:profit}, a point above $y=0$ represents a profit while one below it represents a loss.
The point of intersection of a curve with $y=0$ represents the break-even point of an attack.
The amount of BTC spent to perform the attack at this time is equal to the revenue returned.
By analyzing the figure, we attain the following findings:

\begin{enumerate}[wide=0pt, listparindent=3pt,parsep=3pt]
\item For any value $q$ at $t=0$, the attacker turns a profit of exactly $v$ BTC.
Recall in Fig. \ref{Fig:confidence_graph}, for any value $q$ at $n=0$ (or $t=0$), $P_s=1$.
The merchant accepts an unconfirmed transaction giving the attacker a theoretically perfect chance to succeed.
In this example, the attacker is trying to double-spend $v=5$ BTC resulting in a $\mathsf{profit}=5$ BTC for all values $q$ at $t=0$. 

\item When the merchant waits for $n$ confirmations before accepting a transaction, the attacker is forced to mine blocks in order to create a fork in the blockchain and succeed in the attack.
As discussed in Fig. \ref{Fig:confidence_graph}, the probability of success is based on the computational power $q$ of the attackers. 
Larger values of $q$ correspond to higher probabilities of success $P_s$, hence larger profits/losses.
We also know that $P_s$ declines as $n$ (or $t$) increases for all values $q<0.5$.
As a result, the profits eventually turn into a loss as time progresses.
An attacker with a smaller value $q$ begins losing at an earlier time during the attack while one with a larger value $q$ can withstand longer periods before losing.
However, as reflected by the figure, the losses of attackers with smaller values $q$ are less and continue to increase slower than those with larger values $q$.
This is due to the fact that the cost of electricity $e_q(t)$ for attackers with larger values $q$ are larger than those with smaller values $q$.

\item In Fig~\ref{Fig:profit}, an attacker with $q=0.1$ represents the scenario that begins at the maximum possible profit, then continues to decline till the break-even point.
That time is enough for only three blocks to be added to the blockchain.
In other words, three blocks of confirmation for a transaction worth of 5 BTC should give the merchant enough confidence that an attacker with $q=0.1$ will not be able to reverse the transaction.
The attacker will most likely fail to turn a profit if unable to beat the honest miners in the mining process before they add three blocks to the blockchain.
If the attacker continues to perform the attack beyond this point, the cost will continue to increase while the revenue declines leading to a loss.
As a result, the attacker would most likely surrender at the break-even point to minimize any losses.

\item For $q=0.2$ to $q=0.4$, Fig~\ref{Fig:profit} shows that the profit continues to grow as $t$ increases until it reaches a maximum point due to the accumulation of the mining rewards.
Once the chance of success $P_s$ starts to decline with $t$, the profit also begins to decrease until it reaches the break-even point and later turns into a loss.
However, for $q\geq0.5$, the attacker always succeeds.
The profit is represented as a straight line with a positive slope where the slope represents the rate of turning a profit.
\end{enumerate}

In summary, an attacker with a computational power $q<0.5$ will eventually lose at some point as $t$ increases.
On the other hand, an attacker with computational power $q \geq 0.5$ will always succeed with a profit.
However, it is important to note that this analysis does not include the \textit{luck} factor.
Consider two miners with computational powers $q_1$ and $q_2$ respectively, where $q_1 > q_2$.
The miner with computational power $q_1$ has more resources to solve the proof-of-work, therefore can perform mining faster than the miner with computational power $q_2$.
However, the miner with computational power $q_2$ could still find the solution to the proof-of-work first due to the randomness of the exhaustive search performed.
From a probabilistic standpoint, the chances are low.

\section{Bitcoin Network Security} \label{Sec:networksecurity}

Bitcoin is designed to operate over a P2P network.
It is vulnerable to the decentralized network attacks which can escalate other issues.
In this section, we will discuss major network attacks that can compromise Bitcoin and present network related issues.  We also suggest possible countermeasures.

\subsection{Denial of Service Attacks}
Denial of Service (DoS) attacks flood the network with bogus traffic in order to disrupt legitimate services and participating components connected to the Bitcoin network.
As an example, DoS attacks on a mining pool can result in eliminating the pool from the mining competition, hence giving an advantage to other miners.
They could also facilitate double-spending attacks by preventing certain miners from observing the actual transaction flow~\cite{johnson2014game,vasek2014empirical}.

Some nodes prefer to privately connect to the Bitcoin network in order to limit the possibilities of becoming victims of DoS attacks.
However, this limits the nodes to at most 8 outgoing connections.
As the number of private nodes increases in the network, the random topology connection weakens.
With fewer connections between the nodes, information is flooded at slower rates.
There is also no guarantee of the legitimacy of the 8 outgoing connections each private node connects to.
This means that even a private node can still be vulnerable to a DoS attack if it unluckily connects to malicious nodes.

Bitcoin developers are continuously updating the Bitcoin implementation in an effort to minimize the chances of DoS occurrences.
The newer versions analyze the network connections more closely to try to eliminate suspicious nodes from connecting.
Developers also strive to limit certain transactions/blocks from being flooded throughout the network.
New transactions/blocks are given priority over less important ones such as orphan transactions/blocks.
Certain parameters such as block size are also continuously being altered to adjust the network based on its needs. 
However, the nature of the P2P network makes Bitcoin vulnerable to these attacks.

\subsection{Sybil Attacks}

Peer-to-peer networks are also vulnerable to Sybil attacks~\cite{douceur2002sybil}.
In Sybil attacks, the attacker sets up multiple pseudonymous identities from a single node.
In this way, the attacker can acquire an unfair number of shares of the network IP addresses.
The honest nodes in the network can easily be deceived into believing that the IP addresses belong to different nodes.
With a large number of IP addresses, the attacker can monopolize other connections of nodes and control data propagating to them.

A countermeasure to this attack was proposed in the original Bitcoin white paper~\cite{Nakamoto08Bitcoin}.
This countermeasure also presented a solution to the majority decision-making problem.
It is more convenient to have a one-to-one relationship between a computing machine (node) and a vote instead of having one between an IP address and a vote.
An attacker reproducing multiple IP addresses from a single node can no longer make use of them.
Every node must engage in a \textit{proof-of-work} procedure to prove its legitimacy as discussed in Section \ref{subsec:bitcoin_mining}.

Other countermeasures have been taken by the Bitcoin developers to limit Sybil attacks.
Each outbound connection is limited to a single IP address per subnet mask 255.255.0.0 (i.e. x.y.0.0/16).
In other words, a malicious node can theoretically generate 65536 IP addresses per network prefix consisting of 16 bits where only one can be utilized in a requested outbound connection. 
Today, owning a machine with different network prefixes that consist of 16 bits which can generate numerous IP addresses is impracticable.
Malicious users with IP addresses belonging to different network prefixes need to collude in order to pull off such an attack.

Developers can continue to increase the security by limiting outbound connections to larger subnet masks (for example x.y.z.0/24), however, this would limit the connection possibilities to the outbound connections which contradicts the P2P network.
To optimize security, the subnet mask should be modified dynamically based on the available network prefixes of the nodes connected to the network.
This optimization is very challenging since there is no fixed pattern to how or when nodes connect to the network.
In general, this practice is a weak security countermeasure and can slightly increase the security if optimized.

Users should also realize that the majority of node connections are inbound connections (117).
If we were to assume that all the 8 outbound connections of a node are legitimate, there is no guarantee that the inbound connections are genuine.
A private node relies only on its outbound connections to limit its network connections and the data it receives.

\subsection{Eclipse Attacks}
The Eclipse attack on Bitcoin was proposed in~\cite{heilman2015eclipse}. The primary purpose of an eclipse attack as defined originally in~\cite{castro2002secure,singh2006eclipse,sit2002security} is to monopolize all the outbound and inbound connections of a node within a P2P network.
As a result, the victim node becomes isolated from the rest of the network and only receives data fed to it by the attacker. 
By monopolizing the connections of a node, the attacker can control the blockchain view of this node.
The eclipse attack targets nodes that are possibly discoverable; nodes with public IP addresses.
It strives to populate the tried and new tables of nodes with bogus IP addresses by frequently sending the victim nodes unsolicited $\addr$ messages.
When the tables of nodes are full, they begin evicting random IP addresses to replace them with the newer ones.

The attack requires the victim node to restart all of its connections.
Examples that may cause connection restarts include Internet Service Provider outages, power failures or system/software updates.
When the node tries reconnecting to its 8 permitted outbound connections, it will choose the compromised addresses in either the new or tried tables with a bias towards the newest stored IP addresses.
The optimum time to perform the attack is after populating the tables of the victim node with a decent number of controlled IP addresses.
The chances of a successful attack are based on the percentage of the controlled IP addresses and the time an attacker spends performing the attack.

To limit an eclipse attack, some countermeasures have been proposed~\cite{heilman2015eclipse}.
When replacing IP addresses as newer ones arrive, a deterministic eviction method could be used instead of the random eviction technique.
In this way, each IP address is mapped to exactly one slot in the tables rather than multiple slots, requiring the attacker to possess a large number of addresses.
Also, allowing random selection of IP addresses rather than choosing the most recent ones when initiating an outbound connection makes the attack less biased to the bogus addresses of the attacker.
Other measures include checking an evicted IP address before replacing it with a new one.
If the address still connects successfully, there is no reason to evict and replace it with another one.
Feeler and anchor connections are also good methods that can disrupt an attacker.
Other measures such as increasing the size of the tables, allowing more outgoing connections, or banning unsolicited $\addr$ can also greatly limit eclipse attacks.

\subsection{Routing Attacks}
The main purpose of a routing attack is to intercept the network transmitted messages and tamper with them.
The work presented in~\cite{apostolaki2016hijacking} proposed a routing attack on Bitcoin via the Internet infrastructure.
The Border Gateway Protocol (BGP)~\cite{rekhter2005border} is the most widely used protocol when transmitting data between Autonomous Systems (ASs).
An AS manages a set of nodes with similar IP address prefixes and is responsible for routing data between its nodes and other ASs.

The proposed attack intercepts traffic between ASs by performing two independent attacks: \textit{partitioning} attack and \textit{delay} attack.
The attack takes advantage of the fact that ASs do not validate the newly announced BGP routes which could result in possible BGP hijacks.
A malicious AS can announce forged IP address prefixes to deceive other ASs into believing false routing information.
As a result, a successful attacker will be able to intercept all the traffic for nodes with a certain IP address prefix before it reaches its original destination.

The partitioning attack strives to partition the Bitcoin network into two disjoint groups.
One group represents the set of isolated nodes while the other group represents the remaining network.
The attacker, usually an AS, requires BGP hijacking of other ASs.
Once hijacked, the attacker can intercept all inbound and outbound traffic of all the victim ASs.
However, the attacker cannot intercept the traffic of stealth connections.
Such connections include intra-AS, node connections within the same AS, intra-pool, node connections between gateways belonging to the same mining pool or pool-to-pool, private connections established between pools.
Stealth connections can leak data to the isolated group of nodes and result in an attack failure.
Therefore, the attacker must detect such nodes and remove them from the lists of nodes to be isolated.

Once the network is divided into two groups, the attacker can perform the delay attack.
The main goal of the delay attack is to tamper with data propagating to its destination and cause a stall.
The success of the attack relies on the fact that message exchanging ($\inv$, $\getdata$, and $\txx$) is not encrypted. 
If the attacker intercepts the flow of traffic between ASs, it is possible to tamper with these messages without any node learning about it.
For example, when a node within an AS requests data from its peer within another AS, the attacker will intercept the requested message ($\getdata$) and modify the request.
As a result, the sending node will send undesired data and cause the receiving node to resend a request message.
As long as the attack occurs in a 20-minute time frame, the nodes will not lose their connection and will not be aware that their messages are being tampered with.

The authors in~\cite{apostolaki2016hijacking} suggest some countermeasures to limit the routing attacks.
Simple measures include increasing and diversifying AS connections.
Also, monitoring the network information such as round-trip time can help identify potential threats.
More complex measures can include, encrypting messages, using different channels and ports, and simultaneously requesting data from more than one peer.
However, implementing such more complex measures could introduce additional cost and delay.

\section{Bitcoin Storage Security} \label{Sec:storagesecurity}

Unlike physical wallets that are used to hold cash and banking cards, Bitcoin wallets behave differently. A bitcoin wallet does not store actual BTC.
Instead, it stores the private and public key pairs that can be utilized to prove the ownership rights to certain BTC stored over the blockchain.
As discussed in Section \ref{subsec:transaction}, keys are generated using pseudo-random number generators and elliptic curve cryptography.
In this section, we discuss the variations in Bitcoin wallets and outline the security issues in each.

Similar to~\cite{antonopoulos2014mastering}, we first discuss Bitcoin wallet security based on the key generation and infrastructure of the wallet. Three types of wallets have been defined in BTC.
We summarize the comparison between all three types in Table~\ref{tab:infra_comp}.

The simpler wallets are categorized as nondeterministic wallets, sometimes referred to as \textit{Type-0 wallets}.
In these wallets, when a new pair of keys is requested, the wallet generates a random private key as shown in equation~\eqref{eq:privatekey}. 
Next, the wallet derives its corresponding public key as described in equation~\eqref{eq:publickey}.
The generated key pair is completely random and uncorrelated to the previously generated keys.
However, these wallets require sophisticated management and could fail to perform well as the number of stored keys grows exceeding the storage capacity of the wallet.
A consistent backup of the generated keys is also essential to ensure that the users can still access their BTC in the event of a wallet being unavailable.
However, backups are liable to theft and can result in exposing all the keys belonging to a wallet.

Deterministic wallets are another type of BTC wallets.  
They are also referred to as \textit{Type-1 wallets} and can handle the drawbacks of type-0 wallets.
In this type of wallet, all the generated keys are based on a common and randomly chosen seed $s$.
Using the $s$, all the keys are derived in a deterministic manner.
First, a private key with an index $i$ is generated as
\begin{equation}
\Pr_i = \SHA(s || i) \label{eq:privatekey2}.
\end{equation}
Using equation \eqref{eq:privatekey2}, the corresponding public key $\Pub_i$ is then generated as discussed in equation \eqref{eq:publickey}.
In contrast to nondeterministic wallets, deterministic wallets need only to keep a backup of $s$ to regenerate all of the previously derived keys.

\begin{figure}[t]
\centering
\includegraphics[scale=0.5]{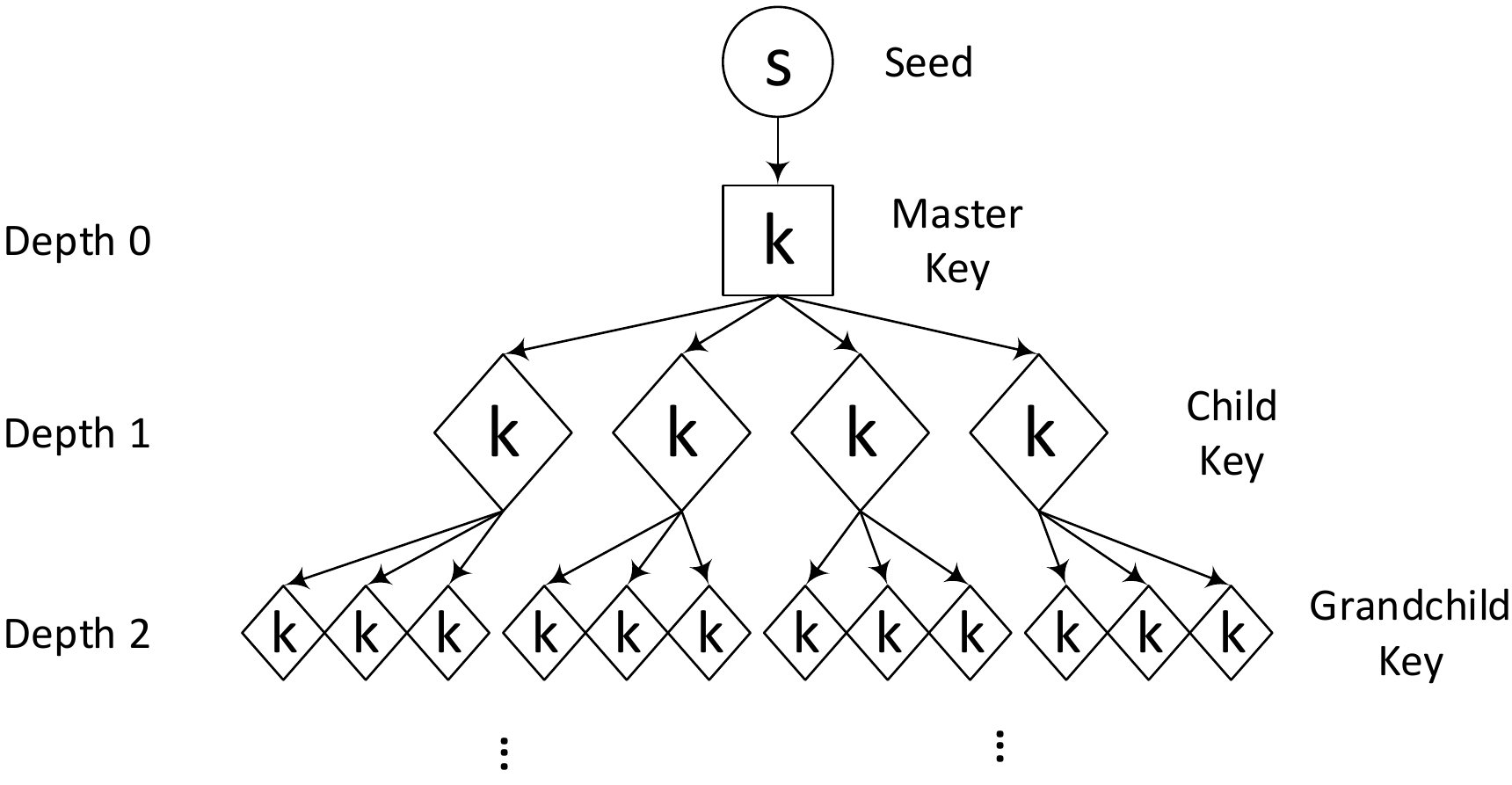}
\caption{The structure of a hierarchical deterministic wallet.}
\label{Fig:hdwallet}
\end{figure}

Hierarchical Deterministic (HD) wallets, referred to as \textit{Type-2 wallets}, were later introduced based on the $\bip0032$ standard~\cite{bip32}.
In HD wallets, keys are generated in a tree structure as shown in Fig. \ref{Fig:hdwallet}.
The key of a node is generated using its corresponding parent node key.

For each node, a key consists of three components: a private key $\Pr$, a public key $\Pub$ and a chain code ($\CC$).
The chain code is a third component introduced to prevent the derivation of the key of a child node from only the private and public keys of the parent node.
In this way, the \textit{extended key} is an extension of both the private and public key.
The \textit{extended private key} is a combination of the private key and chain code which is used to derive the private key of a child node.
Using the derived private key of the child node, it is possible to derive its corresponding public key as explained in equation~\eqref{eq:publickey}.
On the other hand, the \textit{extended public key} is a combination of the public key and chain code which is used to derive the public key of a child node.
It is important to realize that the public key of a child node can be derived using either the extended private or extended public keys.

Key generation begins at depth 0 which derives the root node (master) key components using a randomly chosen seed ($s$).
In many wallets, $s$ is in the form of a mnemonic word sequence as described in $\bip0039$ standard~\cite{bip39}.
A mnemonic word sequence is a sequence of English words that represents a random number used to derive $s$.
Using $s$, the master private key $\Pr^M$ and chain code $\CC^M$ are derived as
\begin{align}
\Pr^M &= \leftbits(\HMAC(s)) \label{eq:masterkey},\\
\CC^M &= \rightbits(\HMAC(s)),
\end{align}
where $\HMAC$ is a one-way hash-based message authentication code that outputs a 512 bit digest and functions $\leftbits$ and $\rightbits$ extract the left and right 256 bits of the digest respectively.
Using the result in equation~\eqref{eq:masterkey}, the master public key $\Pub^M$ is generated as described in equation~\eqref{eq:publickey}.

\setlength{\tabcolsep}{5pt}

\begin{table}[h]
\centering
\caption{Wallet Infrastructure Comparison.} 
\label{tab:infra_comp}
 \begin{tabular}{|c|c|c|c|c|}  \hline
 \textbf{Infrastructure} & \textbf{Management} & \textbf{Seed} & \textbf{Backup} & \textbf{Structure} \\ \hline
 \textbf{Type 0} & Complex & No & All keys &  Random\\  \hline
 \textbf{Type 1} & Moderate & Yes & Only the seed & Sequential \\ \hline  
 \textbf{Type 2} & Simple & Yes & Only the seed & Hierarchical \\  \hline
\end{tabular}
\end{table}

The next step is to generate keys for the children nodes at depth 1 in the tree.
Keys can be generated differently depending on the security of the environment in which the wallet is being used.
For example, when used in a secure environment, the wallet uses the extended private key to generate all the components of a child node key.
This includes the private key which would allow the user to spend BTC from the wallet.
Using the private key $\Pr^p$ of the parent, we can generate the corresponding public key $\Pub^p$ and derive both the private key $\Pr^c$ and chain code $\CC^c$ of the child using a Child Key Derivation (CKD) function as 
\begin{align}
\Pr^c \!&=\! \leftbits(\HMAC(\Pub^p||\CC^p||i)) + \Pr^p  \label{eq:child_pr},\\
\CC^c \!&=\! \rightbits(\HMAC(\Pub^p||\CC^p||i))  \label{eq:child_cc1},
\end{align}
where $\Pub^p = \Pub^M$, $\Pr^p=\Pr^M$, $\CC^p=\CC^M$ are the public key, private key, and chain code of the parent node respectively and $i$ is the index of the child node.
Using the result of equation~\eqref{eq:child_pr}, we can derive the public key $\Pub^c$ of the child node as explained in equation~\eqref{eq:publickey}.

On the other hand, when used in an insecure environment, the wallet uses the extended public key to derive only the public key and chain code of the child node instead of the private key.
This protects the private key from being exposed to potential attackers. 
It also allows payments to be made to the wallet while preventing them from being spent. 
The public key $\Pub^c$ and chain code $\CC^c$ of the child node are derived using the CKD function as
\begin{align}
\!\!\!\Pub^c \!\!&=\! \leftbits(\HMAC(\Pub^p||\CC^p||i)) \!+\! \Pub^p \label{eq:child_pub}, \\
\!\!\!\CC^c \!\!&=\! \rightbits(\HMAC(\Pub^p||\CC^p||i)) \label{eq:child_cc2}.
\end{align}

Although using the extended public key is more secure as it does not expose the private key, it may still put the wallet at risk.
The extended public key exposes the chain code which is an essential component in key derivation.
Using an exposed chain code and public key, an attacker can perform a brute-force attack on all the chain codes derived from it as shown in equation~\eqref{eq:child_cc2}.
In other cases, if the private key of a node is compromised in any way, the attacker can use it along with its corresponding exposed chain code to derive the extended private keys of all the descending children nodes as shown in equations~\eqref{eq:child_pr} and \eqref{eq:child_cc1}.
We also consider the worst case scenario where an attacker is capable of reversing a derived $\Pr^c$ as shown in equation~\eqref{eq:child_pr}.
If successful, using the corresponding parent extended public key, an attacker can derive $\Pr^p$.

To counter these issues, HD wallets also implement an enhanced derivation function known as the \textit{hardened} CKD. 
This derivation strives to secure the exposed chain code within an extended public key.
It prevents the public key of a child node from being derived from the extended public key.
Therefore, the extended private key of the parent node is only useful to derive a hardened private key $\Pr_h^c$ and chain code $\CC_h^c$ of the child node as
\begin{align}
\Pr_h^c &= \leftbits(\HMAC(\Pr^p||\CC^p||i)) + \Pr^p \label{eq:child_pr_h},\\
\CC_h^c &= \rightbits(\HMAC(\Pr^p||\CC^p||i)).
\end{align}
Using the result in equation~\eqref{eq:child_pr_h}, the corresponding hardened public key $\Pub_h^c$ of the child node can be derived as explained in equation~\eqref{eq:publickey}.
In practice, it is suggested to derive the children keys of the master node using the hardened CKD to keep the master key as secure as possible.

Bitcoin wallets can also take other measures to increase the security of storing keys.
Practices such as P2SH~\cite{bip16} and Multi-Sig transactions increase the security of the BTC stored in the wallet, as discussed in Section \ref{subsec:transactiontypes}.
Such techniques are referred to as threshold techniques as they require $M$-of-$N$ private keys to enable BTC spending.
Other wallets enhance the security by encrypting the stored private keys along with a pass-phrase chosen by the owner of the wallet as defined in $\bip0038$ standard~\cite{bip38}. That is
\begin{equation}
\Encrypt(\Pr) = \aes_k(\mathsf{Pr || PassPhrase}),
\end{equation}
where $\aes$ is the Advanced Encryption Standard~\cite{daemen2013design} and $k$ is the encryption key.
If a user wishes to spend BTC, the user must first decrypt the corresponding encrypted private key using $k$ and the pass-phrase previously used in encryption.
Although encryption provides higher levels of security, the user must keep the pass-phrase and encryption keys stored securely.
\setlength{\tabcolsep}{3pt}
\begin{table}[h]
\centering
\caption{Wallet Function Comparison.}
\label{tab:func_comp}
 \begin{tabular}{|m{1in}|m{1.6in}|m{1in}|} \hline
 \textbf{Type} & \textbf{Function} & \textbf{Examples} \\ \hline
 \textbf{Full Service} (online) & Generate private key, derive public key, distribute public key, monitor output TX, create/sign unsigned TX, broadcast TX & coinbase.com~\cite{coinbase.com}, blockchain.info \cite{blockchain.info}\\ \hline
 \textbf{Signing-only} (offline) & Generate parent private key, derive parent public key, sign TX & Ledger Nano \cite{ledger}, TREZOR \cite{trezor} \\ \hline  
 \textbf{Distributed} (offline) & Derive CKD, Distribute public key & customized pre-populated database \\  \hline
\end{tabular}
\label{tab:wallets}
\end{table} 

The wallets that exist today come in different forms and account for different security measures.
Based on the different installation environments, wallets can be categorized into three types: online (web) wallets, desktop (software) wallets, and mobile wallets.
As in~\cite{wallets}, we can further categorize each type of these wallets into: full-service wallets, signing-only wallets and distributing wallets, based on the functions that they can perform.
Table~\ref{tab:wallets} summarizes these different functions.

A full-service wallet is one that can perform all the functions required to spend and receive BTC.
These functions include generating private keys needed to spend BTC, signing transactions with the private keys, deriving public keys needed to receive payments of BTC, broadcasting the derived public keys to the network, and monitoring the BTC spending and receiving of a wallet.
Full-service wallets must be able to connect to the Bitcoin network. 
Examples of online full-service wallets include the wallets provided by coinbase.com~\cite{coinbase.com} and blockchain.info~\cite{blockchain.info}.
Armory, Electrum and Bitcoin Core are the most popular desktop full-service wallets today.
For mobile wallets, an example that runs on both Android and iOS includes the Airbitz wallet.

The second type of wallets are the signing-only wallets. 
The main purpose behind these wallets is to enhance the security of the wallet by generating private keys in secure offline environments.
Working in conjunction with a networked wallet, the signing-only wallet can interact with the Bitcoin network and can deterministically generate pairs of private and public keys as needed to transfer the public key to the networked wallet.
The role of the networked wallet is to distribute the public key to allow payments to be made to the wallet.
In case of an HD wallet, the network can also generate child node keys as desired.
Once the networked wallet detects a transaction addressed to one of the public keys that it has distributed, it creates an unsigned transaction based on the $\utxo$ and transfers it to the signing-only wallet.
The signing-only wallet then uses its private key that could be derived from an extended private key in the case of an HD wallet to sign the transaction and returns it back to the networked wallet.
Finally, the networked wallet distributes the signed transaction in the Bitcoin network to claim the BTC.

Signing wallets can either be offline wallets or hardware wallets.
Offline wallets are designed to reduce the network vulnerabilities.
Their tasks include private key derivation and transaction signing.
The signed transactions are transferred via removable media to the online wallets.
Offline wallets provide higher levels of security than the full-service wallets, however, they require a continuously isolated device.
On the other hand, hardware wallets are less of a hassle than offline wallets.
They are connected directly to the networked device which eliminates the dependency of removable media when communicating between the signing-only wallet and the networked model.
However, the hardware wallet is also inconvenient in situations where the owner makes frequent payments since the owner must constantly carry the hardware wallet to be able to make a payment anytime. 
As a result, many people use hardware wallets for long-term storage rather than day-to-day transactions.
Utilizing this type of wallet, one can store large amounts of BTC in the most secure environments.
Popular examples of hardware wallets today include the Ledger~\cite{ledger} Nano and TREZOR~\cite{trezor}.

The final type of wallets are the distributing-only wallets. 
These wallets also strive to reduce the security issues caused by the full-service wallets.
They are in the form of networked wallets for public key distribution in a pre-populated manner, where the public keys are derived and distributed as needed by the network.
Other distributing-only wallets are capable of generating the public keys as the case in HD wallets.

Exchange platforms store large portions of cryptocoins in online wallets to provide their users the advantage of reduced transaction time due to the immediate availability of their private keys.
This is analogous to storing cash in a centralized entity such as a bank.
It is important to point out that storing cryptocoins in an online wallet provided by an exchange platform is the least secure method since it means storing the corresponding private keys that can spend those cryptocoins.
The users must completely trust the exchange platform to safely store the private keys and not act maliciously.
Even worse, assuming we can trust an exchange platform, cryptocoin owners are still at risk of losing their cryptocoins in the event the exchange platform online wallets are hacked and the private keys are leaked.
A hacker that gets a hold of the private keys can immediately use them to send the cryptocoins to his/her personal address.
Once the transaction is processed and stored over the blockchain, it becomes immutable to being deleted/modified and most likely will not be reversed unless the blockchain is hard forked.

Throughout the history of cryptocurrencies, multiple attacks have occurred to exchanges that resulted in massive losses and severe price panics to certain cryptocurrencies.
In 2011, one of the most notable Japanese-based exchanges, Mt. Gox, online wallets were hacked, leaking all the private keys it stored in the wallet.dat file.
Mt. Gox was able to recover from that heist, however, later in 2014, it filed for bankruptcy and was shut down since it was responsible for around 70\% of Bitcoin trading volume and lost approximately 850,000 BTC that was valued at more than \$450 million dollars.
The hackers were able to even steal BTC stored in the exchange's hardware wallets.
There is no legitimate evidence of how the attack occurred.
In March 2014, Mt. Gox reported on its website that it had found 200,000 BTC from the total stolen in old-format digital wallets.
The other 650,000 were believed to be laundered on another exchange platform known as BTC-e.

The problem is that such heists could possibly occur again.
Exchange platforms remain to be an extremely attractive hacking points for hackers since they hold so many funds in the least secure manner.
Users are recommended to keep limited amounts stored in exchanges while storing the majority of their funds in hardware wallets.

Another issue is whether or not it is possible to track the movement of stolen cryptocoins, hence, catch the hacker.
Based on our analysis, it is theoretically possible.
However, there have been scenarios where hackers were able to launder large portions of stolen cryptocoins such as the example discussed previously.
Another famous example occurred in January 2018 when about \$534 million dollars worth of a cryptocoin known as XEM were stolen from a Japanese-based exchange known as Coincheck.
Today, this heist represents the largest theft in the history of cryptocurrencies.
The exchange also announced that the cryptocoins were stolen from its online wallets through multiple unauthorized transactions.
In an effort to combat this fiasco, the developer team announced that they will develop an automated tagging system to tag stolen XEM cryptocoins.
However, the tracking system was ineffective. 
Once more, the stolen cryptocoins were laundered and completely lost.

In conclusion, we stress on the fact that there remains to be a trade-off between the security of a wallet and the ease of use.
The most frequently used wallets today are full-service wallets. 
They are free, user-friendly and can perform all functions needed by a BTC owner.
However, these wallets could be vulnerable to theft since they are connected to the network.

\section{Bitcoin Privacy} \label{Sec:privacy}

Bitcoin suffers inherent privacy issues in that attackers could link certain identities to their pseudonyms (such as Bitcoin addresses) and identify their history of transactions. 
This is known as the \textit{linking problem}.
Many users publish their real identities and Bitcoin addresses online so that others can make payments to them.
This practice is common among blogs and websites that request BTC as donations or those selling a product or service.  
These actions could jeopardize their anonymity.
Another common example is when users trade BTC for other altcoins over exchange platforms.
Most exchange platforms require users to validate their identities by uploading a copy of official identification which exposes the users to the exchange applications.
Such examples do not require an attack to learn the full transaction history of those users. 
Simply by tracing the Bitcoin addresses over the blockchain, the transactions could be revealed.
In fact, even cautious users that do not publicly use their identities may be at risk as well.

Bitcoin utilizes Bitcoin addresses as its defense mechanisms to preserve the privacy of users.
When generated for the users, bitcoin addresses do not leak any information about the identities of the users.
However, attackers strive to search for links between bitcoin addresses and user identities using auxiliary information available over the network.
If a link is found, it is possible to discover all the other Bitcoin addresses belonging to that user and revealing the complete history of BTC transactions of the user.
Today, powerful analysis tools and search engines can be utilized to discover the Bitcoin address and determine this information.
Even the strongly encouraged practice of using a new Bitcoin address for every new transaction cannot completely prevent this information from being revealed once a Bitcoin address is linked to an identity of the user. 

The auxiliary information can be obtained by multiple methods.
Different techniques exist today that can speculate links between Bitcoin addresses and user identities.
The study in~\cite{narayanan2009anonymizing} shows that using information about how nodes are connected within a network can help identify users.
In~\cite{crandall2010inferring}, it was shown that patterns of co-occurrences may reveal useful information and lead to any ties.
The study in~\cite{puzis2009collaborative} showed that just by monitoring the communication channel, users are likely to lose their anonymity.
In~\cite{korolova2008link}, an analysis is presented that shows how compromised network nodes can leak significant user information and link them to certain transactions.
For further reading, we direct reader to~\cite{altshuler2013stealing,reid2013analysis,meiklejohn2013fistful,androulaki2013evaluating} that present similar studies.

Users can run their nodes over Tor~\cite{dingledine2004tor} in an effort to hide their information from the rest of the network.
Tor is a software that provides an additional layer of anonymity.
It utilizes multi-layer encryption and random relaying nodes to transfer data between a sender and receiver.
The sender begins by sending the multi-layer encrypted message to a random node that decrypts a single layer and transmits it to the next relaying node.
This process continues until the message is completely decrypted and arrives at the receiver~\cite{RenWu10Survey}.
However, multiple studies, such as~\cite{biryukov2014deanonymisation,biryukov2015bitcoin,overlier2006locating,dingledine2014one}, have shown that even a low-resource attacker could be capable of gaining information flowing between users running their Bitcoin nodes over Tor.
This information can include the data sent between nodes or even the location of the nodes within the network topology.

Other efforts have also been employed in an effort to improve the anonymity of Bitcoin.
We classify these efforts into two main classes: mixing services and joint transaction.

\subsection{Bitcoin Mixing Services} \label{subsec:btc_mixing}

BTC mixing is an approach that mixes identifiable BTC in an effort to make them unrecognizable by public observers.
The first generation mixing was centralized and performed by \textit{tumblers}.
Tumblers are third party mixers that receive BTC from different users, randomly mix them up, and then return to the users their updated BTC amounts.
An attacker would no longer be able to trace the BTC of a certain user since the user no longer possesses the same BTC that he/she previously owned.
However, a tumbler being a centralized entity presents many threats to the users.
It must be fully trusted not to steal the BTC it mixes or even leak any information about the mixing process.
Even when completely trusted, being centralized makes it prone to being compromised.
In addition to this, tumblers charge users mixing fees in return for their services.

In an effort to mitigate these risks, a new generation of peer-to-peer tumblers was introduced to decentralize the process.
Instead of sending BTC to a tumbler that performs mixing, the users themselves are involved in the process.
This eliminates the need to completely trust a third party and minimizes the risk of privacy leakage.
An example of such a protocol is CoinSwap which is presented in~\cite{coinswap}.

\subsection{Bitcoin Joint Transactions}

A joint transaction allows different users to combine the inputs and outputs of their transactions into a single transaction to be processed as a whole.
All participating users must provide their own signatures to the transaction to unlock their input portion.
Once all participating users correctly sign their inputs, the transaction can be processed as a regular transaction and added to the blockchain.
An attacker can no longer trace the BTC movement of a user since there is no direct relationship between the inputs and the outputs of a transaction.
The level of privacy provided by a joint transaction increases with the number of participating users.
This also results in a lower transaction fee that is paid by each user as it is divided among more users.
In 2013, Gregory Maxwell introduced this concept as CoinJoin~\cite{maxwell2013coinjoin} which is widely used in practice today.
CoinJoin eventually began to evolve and existed in multiple flavors.
Notable examples that introduced new concepts are described below. 

\para{SharedCoin} \textit{SharedCoin} provided by Blockchain.info is one of the initial implementations of the CoinJoin protocol that ran over a centralized server. 
The centralized server was the meeting room for the participating users to meet and combine their transactions together.
Since users meet in one place, the server is capable of keeping logs of the transactions processed over it.
This requires users to completely trust the server not to misuse these logs and put their information at risk if compromised.
Shortly, Kristov Atlas created CoinJoin Sudoku, a software that is capable of analyzing the mixing process performed by SharedCoin. 
The software aims at discovering the relationships between transactions and their owners.
It clustered matching inputs and outputs of transactions trying to identify a common owner.
However, this implementation is completely suspended today due to its various privacy limitations.

\para{Dark Wallet} In 2013, Cody Willson and Amir Taaki introduced \textit{Dark Wallet}~\cite{darkwallet}.
It provides anonymity using stealth addresses and the CoinJoin protocol. 
A stealth address is a public seed address combined with some metadata used to derive an actual address for a payee to receive transactions. 
The metadata is shared only between the payer and the payee, and cannot be accessed by the public observers.
To generate an actual address, the payee generates a private key and its corresponding public key.
Next, the payer uses the public key of the payee and some metadata to generate a transaction with a new address.
Once the payee learns the metadata, it can claim the amount attached to the transaction by deriving the appropriate key from the stealth address.
Others trying to trace the transaction that was received with a stealth address would not be able to trace it.
However, Dark Wallet cannot provide complete anonymity against linking users to certain BTC transactions since the payer can trace it.

\para{CoinShuffle} \textit{CoinShuffle} was introduced in 2014~\cite{ruffing2014coinshuffle}. 
It is a combination of the CoinJoin protocol and the accountable anonymous group communication protocol Dissent~\cite{corrigan2010dissent}.
Its main purpose is to eliminate the involvement of third parties while achieving anonymity and protection against DoS attacks.
The protocol consists of three main phases: announcement, shuffling and transaction verification.
In the announcement phase, the participants generate a new pair of private and public keys then broadcast their corresponding public key to the other participants.
In the shuffling phase, each participant generates a new Bitcoin address to be used as their output address in the mixing transaction.
Following that, the participants obliviously shuffle these generated Bitcoin addresses.
In the transaction verification phase, every participant checks whether their Bitcoin address is contained in the output list.
If present, each participant creates a mixing transaction that spends the inputs to the shuffled list of outputs, signs the transaction, and broadcasts the signature.
Once each participant receives the signatures of the others, every participant can generate a fully signed version of the mixing transaction.
Dishonest behavior can be detected by the presence of one honest participant who would not broadcast his/her signature and report the dishonesty to all other participants.

However, Coinshuffle suffers anonymity vulnerability if not used cautiously since it allows users to assign change back to themselves in the mixing transaction.
Once the change is assigned to the Bitcoin address of the user, anonymity could easily be lost.
The best solution to this problem is to use amounts that do not require any change.
However, the user does not necessarily get to choose what amount to use since the user must use $\utxo$(s) from previous transactions.
In addition to this, Coinshuffle reveals the identities of the participants among each other during the process.

\para{JoinMarket} \textit{JoinMarket}~\cite{joinmarket} is a decentralized CoinJoin implementation. 
It aimed at improving the privacy of all the previous implementations.
JoinMarket introduced two types of participating users, market makers, and market takers.
Market makers are users who are willing to mix their BTC at any given time in return for a fee.
On the other hand, market takers are users that demand immediate mixing service and are willing to pay a fee as compensation to the market makers.
Market makers and takers negotiate the service over an Internet Relay Chat (IRC) channel.
Once terms are discussed, a mixing contract is generated which enables each participating user to operate from their own personal machine.
The fact that the system is decentralized protects users from the need to trust a centralized entity.
Furthermore, the fee paid by the takers to the makers incentivizes them to continue to join.

Various protocols continue to evolve in an effort to increase the anonymity of Bitcoin.
However, the linkage problem still remains within Bitcoin that could jeopardize the anonymity of its users.

\section{Security and Privacy of Altcoins} \label{Sec:altcoins}

The continuous emergence of altcoins presents enhanced features to the cryptocurrency enthusiasts.
Some of these altcoins have proven to provide enhanced security and privacy over Bitcoin.
However, Bitcoin continues to remain at the top of the list of cryptocurrencies with the largest market cap.
This contradiction raises questions around its continuous dominance.

In this section, we unfold the major security and privacy advantages of altcoins.
We first investigate distinct consensus algorithms implemented by different altcoins in an effort to keep their network secure.
We strive to elucidate the security advantages of these algorithms over the proof-of-work implemented by Bitcoin.
Next, we discuss major altcoin privacy protocols and privacy improvement over Bitcoin.

\subsection{Altcoin Security}
The Proof-of-Work (PoW) implemented in Bitcoin utilizes $\SHA$; a CPU-bound function. 
The time needed to run $\SHA$ is determined by the speed of the machine.
Powerful machines such as ASICs can run $\SHA$ millions of times faster than various other CPUs, GPUs, and FPGAs.
This created an unfair mining competition since not all miners use the same computing machine.
In fact, it eliminated miners using CPUs, GPUs, and FPGAs since their chances of success are negligible when compared to those using ASICs.

This PoW has also been greatly criticized for being an energy-wasting technique.
Mining is performed using powerful computing machines that require substantial energy to run.
Most of the energy used by all these miners ends up being wasted since the output of only one miner is used to extend the blockchain.
As a result, the cost of running this PoW to achieve consensus is extremely costly.

In addition to this, it is expected that Bitcoin will suffer a mining tragedy of the commons~\cite{hardin2009tragedy}.
The mining reward will converge to zero since it continues to halve approximately every four years (precisely every 210,000 blocks). Eventually, the miners will no longer have an incentive in taking part of the consensus procedure someday.
This will force the transacting users to increase their transaction fees as an alternative incentive to the miners.
As a result, both the users and miners will be driven away from the system.

In an effort to mitigate these issues, some altcoins replaced $\SHA$ with memory-bound hash functions in their PoW.
In comparison to the CPU-bound function used by Bitcoin, the time needed to run memory-bound functions is determined by the amount of memory available to hold the processed data.
Developing ASICs for memory-bound functions is no longer advantageous since they can only optimize CPU-bound functions.
Notable examples of such functions include scrypt~\cite{percival2009stronger} and CryptoNight~\cite{saberhagen2013crypto}
Combinations of hashing algorithms have even been used such as X11~\cite{x11} and X12-X17~\cite{xs}. 
In these algorithms, 11-17 different hashing algorithms used. 
The result of each sub-algorithm is fed as input to the next sub-algorithm. 
Popular altcoins  such as Litecoin, DASH~\cite{duffield2014dash}, and Monero implement such examples.
However, it was not too long until optimized memory-bound ASICs started re-monopolizing the mining process once again.

Developing an ASIC-resistant PoW has not succeeded.
Altcoin developers began to deviate their efforts to implement alternative consensus algorithms that strive to mitigate ASIC centralization and prevent critical issues such as double-spending attacks.
Similar to PoW, many of these alternative consensus protocols are chain-based.
A chain-based protocol pseudo-randomly selects a single validator to generate the next block of the blockchain.
Some widely implemented consensus chain-based protocols are described below.

\para{Proof-of-Stake (PoS)} PoS is an alternative consensus algorithm that was initially suggested in~\cite{pos}.
In contrast to PoW, PoS is dependent on economic stakes of users (i.e. holdings in cryptocurrency) rather than their computational resources.
The algorithm deterministically selects a user with significant holdings to validate the next block. 
In return, the selected validator is rewarded a certain value of the cryptocurrency similar to the mining reward in PoW and all the transaction fees included in the block.
Conceptually, a user holding $x\%$ of the total available cryptocurrency will be chosen $x\%$ of the time as the validator in generating the next block.
Once the block is generated, the validator relays it to the other validators to confirm it and extend the blockchain.

PoS has multiple benefits in comparison to PoW.
Users are no longer required to consume substantial quantities of electricity since they no longer engage in a mining process.
In fact, they are motivated to take part in the validation process as it requires nothing more than presenting their wealth in return for a reward if chosen to be the validator. 
In contrast to PoW, PoS significantly speeds up the consensus process.
From a security perspective, PoS tackles the 51\% attack by making it more expensive than performing it in a PoW environment.
An attacker would need to possess 51\% of the total cryptocurrency available to perform the majority attack.
Assuming a single user possesses 51\% of the total cryptocurrency and performs the attack, the value of the cryptocurrency will drop and the attacker would suffer most being the majority stakeholder.
In comparison to PoW, the majority attack requires 51\% of the total mining power which is theoretically achievable through mining pools.
This incident previously occurred in the mining environment of Bitcoin as a mining pool (Ghash.IO) exceeded the 51\% threshold.

Although PoS could handle some issues caused by PoW, it also introduced some major challenges.
The largest stakeholders will be able to monopolize the consensus procedure as they will always be selected and earn the reward. 
This will create a centralized consensus environment.
In addition to this, an attacker with a 51\% stake can also completely destroy the cryptocurrency, assuming the intentions of the attacker are to eradicate the system at any price.
PoS also suffers a major flaw known as Nothing at Stake (NoS).
This issue can occur if coincidentally two stakeholders are chosen to validate the next block.
This may result in two valid blocks that can extend the blockchain.
As a result, a fork may occur to the blockchain as the miners accept both blocks.
To resolve the fork, the validators vote on both branches.
Voting is done at no cost which may be an incentive for a malicious validator to vote for a specific path of the blockchain and facilitate a double-spending attack.

These issues resulted in PoS to start appearing in multiple flavors.
Its first implementation appeared in a Bitcoin fork, namely Peercoin, which incorporates a hybrid of an energy-efficient PoS~\cite{king2012ppcoin} and the original PoW that runs $\SHA$. 
PoW was used initially as a method of coin generation and distribution to get the system running.
As time progressed, PoW was slowly replaced by PoS to validate transactions, mint new coins and maintain consensus.
The validators are chosen based on the number of coins in their possession and their corresponding age (i.e. a timestamp indicating how old the coins are).
Once they are granted a reward in return for their service, the age of their coins goes back to zero to give other validators a chance to generate the next block.
By that, no single validator can monopolize the validation process. 

Later, modified versions of PoS were implemented into some cryptocurrencies.
In~\cite{vasin2014blackcoin}, the age of the coins was removed as it was argued to be abusive to the system.
It can help gain significant network weight and facilitate a double-spending attack.  
In some cases, it may also discourage honest users from staking persistently as they would hold back until their coins are oldest in age to maximize their chances.

In~\cite{larimer2014delegated}, a delegated proof-of-stake (DPoS) was proposed where the users vote for validators (referred to as witnesses).
Each vote has a different strength based on the stake of the user.
However, this requires users to completely trust the validators they vote for.

\para{Proof-of-Activity (PoA)} PoA is a consensus algorithm that combines PoW and PoS into one protocol~\cite{bentov2014proof}. 
Its purpose is to reward only the online participators, thus motivate more miners to remain online in an effort to secure the network.
The protocol is analogous to the lottery where the chances of winning of an individual are based on the number of tickets the individual holds.

In PoA, miners first utilize their computational power to compete in generating an empty block header; one that does not reference any transactions.
A successful miner then immediately broadcasts the resulting hash to the network.
This hash value is used to deterministically derive $N$ pseudo-random stakeholders who are potential miners if found to be online.
The derivation of these $N$ stakeholders is performed by hashing a concatenation of the broadcast hash value, the hash of the previous block, and $N$ fixed suffix values.
The protocol then invokes a subroutine known as \emph{follow-the-satoshi} once for each derived value.
The subroutine finds the block storing a satoshi with the same index as the result.
Next, it inspects the block in which the satoshi was minted and traces its movement up until its last owner.
If online, this owner participates in the next block generation process that extends the blockchain.
Similar to PoS, the more satoshis an individual owns, the more likely that the individual will be selected randomly in this process.

Every stakeholder then checks the validity of the empty block header that was initially broadcast.
Using this value they also check whether they were one of the $N$ selected validators.
The first $N-1$ lucky stakeholders sign the hash of the empty block header with the private key that controls the satoshi derived from \emph{follow-the-satoshi} subroutine.
Next, they broadcast their signature to the network.
The $N^{th}$ stakeholder then generates a wrapped block that extends the empty block header by including the desired transactions to be verified, the $N-1$  signatures, and his/her own signature for this block.
The wrapped block is finally broadcast to the network to extend the blockchain.
The transaction fees that the $N^{th}$ stakeholder collects from the included transactions are shared among the miner and the $N$ participators.

From a security perspective, PoA makes the 51\% attack more difficult than PoW and PoS since a large computational power and a significant stake are both required in PoA.

\para{Proof-of-Burn (PoB)} PoB is an algorithm that achieves consensus by \textit{burning} a portion of a cryptocurrency.
Burning a portion of cryptocurrency means generating a transaction with this portion destined to an inaccessible address by all users.
The concept of burning is analogous to buying expensive computational hardware in PoW.

In general, a miner burns portions of his/her holdings and waits a certain period of time.
This time ensures that it is impractical for an attacker to undo the transaction.
After waiting, the transaction is permanently stored in the blockchain and becomes visible to all observers.
This is proof that the potential miner has invested a portion of his/her holdings and is worthy of being a miner.
Honest miners will burn portions of their holdings that are less than or equivalent to what they can return in the mining process if successful.
In other words, if miners burn more than what they are expected to return in a successful mining process, they will spend more than what they earned, hence a loss.

The potential miners then create candidate blocks in an effort to extend the blockchain.
By referencing their transactions in the blockchain, they can prove that they have burnt some of their holdings earlier, thus become accepted by the community as miners.
The winning block that extends the blockchain is chosen by allocating the miner that has burnt the most after a certain period of time.

From a security perspective, this algorithm can achieve the same security as its predecessor algorithms.
It requires a miner to perform an expensive task (burning) that is easily verified by all other participators observing the blockchain.
Similar to PoS, it saves the miners the hassle of buying hardware to physically perform mining.

PoB is also known for its use in bootstrapping new cryptocurrencies.
A new cryptocurrency can mint its new coins by utilizing PoB.
Rather than releasing mint coins during the mining process as in PoW, a cryptocurrency can be burnt to mint coins from the new cryptocurrency.
For example, a new cryptocurrency can mint coins by burning BTC. 

Whether being used to maintain consensus of a network or bootstrap an emerging cryptocurrency, PoB has been criticized for its permanent coin destruction.  
This is a more critical issue for cryptocurrencies with a limited supply.
The more PoB is utilized, the less the quantity of a cryptocurrency is in circulation.
Such an issue can lead to significant inflation to the value of the cryptocurrency which can result in destruction of the cryptocurrency.

In contrast to the chain-based algorithms discussed previously, some alternatives are based on Byzantine Fault Tolerance (BFT) algorithms.
In these algorithms, consensus on a block is independent of the chain.
The algorithms utilize a multi-round process where every validator sends a \textit{vote} for some specific block during each round.
At the end of this process, the validators reach an agreement on whether to permanently accept a given block.
These protocols could be somewhat more centralized since the validators work together to maintain consensus by handling each block individually.
For further reading, the readers are recommended to review examples such as the ripple consensus protocol~\cite{schwartz2014ripple} and the stellar consensus protocol~\cite{mazieres2015stellar}.

\subsection{Altcoin Privacy}

Privacy is one of the most important issues of cryptocurrencies.
Some notable altcoin privacy protocols are described below.

\para{Zerocoin} 
Zerocoin~\cite{miers2013zerocoin} protocol generates anonymous coins that can be exchanged for other cryptocurrencies (e.g. BTC) when mixing is desired.
Cryptocurrencies that embed the protocol run in parallel with Bitcoin utilizing its blockchain.
A user that wishes to mix BTC purchases this cryptocurrency through a \textit{mint transaction}.
A mint transaction trades the specified amount of the cryptocurrency into its corresponding value in the anonymous coins.
Once bought, the user can convert his/her anonymous coins back into BTC through a \textit{spend transaction}. 
The spend transaction trades the anonymous coins back to the original cryptocurrency.
The mint transaction and the spend transaction are designed to be uncorrelated.
After a spend transaction, the user ends up with a different set of BTC than the ones used in the mint transaction.
In comparison to the mixing techniques of Bitcoin discussed previously, Zerocoin eliminates the need for tumblers.
It relies on a combination of digital commitments, one-way accumulators, zero-knowledge proofs and the existing Bitcoin platform.

In a mint transaction, the user first generates a random serial number $s$ and cryptographically commits (i.e. encrypts) it into a coin $c$ using a randomly chosen key $k$.
The purpose of cryptographically committing the serial number is to hide its value from all the other users while binding it to its owner.
The user then generates a Bitcoin transaction with the appropriate amount to pay for coin $c$ (i.e. the BTC to be mixed) and releases both of them into the network.
The miners place coin $c$ into a one-way accumulator and mine the Bitcoin transaction into the blockchain.
The transaction is not addressed to anyone and its value remains locked in the blockchain until it is redeemed by another user in a spend transaction.
At this point, the user possesses anonymous coins equivalent to the amount of BTC that is to be mixed.

To convert the anonymous coins back to BTC, the user exchanges his/her anonymous coins with other locked BTC of users stored on the blockchain.
The user first provides zero-knowledge proof of his cognizance of coin $c$ in the one-way accumulator.
Next, the user must prove that his key $k$ and serial number $s$ correspond to coin $c$.
The miners then verify the proof and that the serial number $s$ has not been previously spent.
Once verified, the spend transaction is mined into the blockchain granting the user an equivalent amount of BTC as spent in the corresponding mint transaction.
The user ends up with fresh BTC that he/she have never possessed. 

However, Zerocoin protocol has a few limitations.
From the perspective of the system, Bitcoin must be soft-forked to account for the changes of the protocol.
From the perspective of the protocol, the zero-knowledge proof computed generates large signatures that would eventually bloat the blockchain.
The process is also time-consuming and requires more time for transactions to be accepted by the system.
Most importantly, it requires a trusted party to initiate the one-way accumulator.

Shortly after the release of Zerocoin, Zerocash~\cite{sasson2014zerocash} also known as Zcash today, was introduced in an effort to reduce the cost of the zero-knowledge proof.
However, this project did not require a soft fork to Bitcoin as Zerocoin protocol did.
In fact, it was a standalone technology that implemented its own cryptocurrency.
It utilizes a smaller sized zero-knowledge proof known as zk-SNARKs~\cite{ben2014succinct} that consumes less time to compute.

\para{PrivateSend} 
PrivateSend is an altcoin joint transaction protocol~\cite{privatesend} that combines identical inputs from various users into one transaction with multiple outputs.
A user initially reaches out to a random master node requesting mixing specific denominations of a certain amount of coins.
The master node then announces that it is willing to accept other coins of identical quantities and denominations to be mixed into a transaction.
Once the master node receives enough requests, the involved users specify their full list of inputs and outputs they wish to be mixed.
The inputs specify the coins to be mixed while the outputs specify the output addresses of users where they wish to receive the mixed coins.
The master node then puts all inputs and outputs into a joint transaction and sends it to the involved users.
The users validate the transaction and sign their inputs and return it to the master node.
The master node finally broadcasts the transaction to the network which is treated as any other transaction.
However, PrivateSend is not a completely decentralized protocol and can jeopardize the anonymity of the user since it involves a centralized node.

\para{CryptoNote} 
CryptoNote is a privacy-preserving protocol embedded in some cryptocurrency implementations that strive to hide the connection between a sender and receiver from the rest of the network~\cite{saberhagen2013crypto}.
The protocol protects the identity of the sender by utilizing ring signatures~\cite{rivest2001leak} when signing transactions.
The public key of the sender is shuffled with public keys of other senders, giving all keys equal probability of being linked to a transaction.
In this way, an attacker has no way of identifying the private key used during transaction signing, hence identifying the sender. 
It also generates a unique public key for the receiver with each new incoming transaction.
Using random data generated by the sender and the public key of the receiver, a one-time unique pair of private and public keys is generated via Diffie-Hellman key exchange~\cite{diffie1976new}. 
These keys are used to claim the transaction output by the receiver.

Unlike Bitcoin, the blockchains used by the cryptocurrencies running CrypoNote do not reveal the information of transactions, hence improving anonymity.
As a result, verifying transactions becomes a challenge.
To handle this issue, a modified version of the original traceable ring signature~\cite{fujisaki2007traceable} is utilized.
The original scheme makes it possible to trace transactions sent by the same sender if they contain the same tag and are signed by the same private key.
The modified scheme, referred to as one-time ring signature, replaces the tag with a key image.
The key image is deterministically derived by applying a cryptographic hash function to the private key allowing each sender to generate only one valid signature using his/her private key.
If the sender tries to generate two different signatures with the same private key, a link will be detected.
Therefore, this scheme counters any double-spending attempts since the blockchain will only store one signature while invalidating all others.

\section{Conclusions and Future Research directions} \label{Sec:conc}
After presenting our extensive survey, we recap the lessons learned and the future research directions that can be derived from this study.

In this survey, we strived to address major technical concerns regarding the future stability of Bitcoin.
We first introduced the background of Bitcoin and explicated its major building blocks and protocols.
The main purpose of our extensive background was to educate our readers about the blockchain technology using Bitcoin as a use case.

Next, we delved into crucial security concerns.
We began by discussing the double-spending attack and analyzed its probability of success.
We showed that the probability of success can be modeled using two different probabilistic models that result in a similar outcome.
Using this analysis, we further evaluated the profitability of the attack.
We showed that attackers with less than half of the total computational power of the system will eventually lose at some point while performing the attack. 
The main lesson learned was that there will always be a trade-off between the waiting time before accepting a transaction and the possibility of reversing the transaction.
Users should realize this trade-off and only use the current systems acknowledging these risks.

Following that, we also explored the major network-related security issues of the underlying peer-to-peer network.
Our discussion showed that these network attacks are inevitable since it is impossible to restrict malicious nodes from connecting to the network. 
We further explored storage security by investigating the wallet infrastructures and the different modes of storage. 
Our analysis shows that there is also a trade-off between storage security and practicality: the more user-friendly wallets are, the bigger risk of losing their cryptocoins.
Therefore, users must take all precautions in order to protect their funds.

Beyond the security issues that Bitcoin suffers, we investigated some privacy limitations inherent to the system. 
We debunked the misconception of Bitcoin anonymity and reviewed major methods for privacy protection. 
Currently, systems similar to Bitcoin continue to suffer privacy issues.
The privacy of the users is at risk.

Finally, in the last section, we also looked to expand the knowledge of the readers on some emerging protocols that have been implemented in some altcoins. 
The main purpose of these protocols is to enhance security and privacy. 
However, our discussion proved that even these emerging protocols have not been able to provide good enough systems to completely eliminate the current centralized systems.

As blockchain has presented many intriguing features, most importantly decentralization, it has also introduced new research challenges.
Future research must find ways to combat these concerns in order for the stability of blockchain to become consolidated. 
Based on our survey, we encourage future research to expand on our mining profitability analysis in order to help users come to better decisions before utilizing such systems.
We believe that in order for systems such as Bitcoin to attract massive adopters, users must clearly understand the risks and gain a certain degree of confidence.

We also believe that extensive research is required to enhance security and privacy protocols.
Typical peer-to-peer network attacks and privacy concerns such as the ones discussed in this paper can be used to disrupt the stability of blockchain systems.
Currently, all the evolving solutions may enhance the security and/or privacy slightly, but it usually comes with a price keeping the users skeptical about using such systems.
However, as research advances in these fields, cryptocurrencies may help revolutionize the payment system as we know it today.

\bibliographystyle{ieeetr}
\bibliography{Bitcoin}

\end{document}